\documentclass[aps,prd,twocolumn,superscriptaddress,preprintnumbers,floatfix,nofootinbib]{revtex4-2}

\usepackage{graphicx}
\usepackage{dcolumn}
\usepackage{bm}
\usepackage{hyperref}

\usepackage{amsmath}	
\usepackage{multirow}

\usepackage{color}

\usepackage{orcidlink}

\newcommand{\aj}{AJ}			
\newcommand{\araa}{ARA\&A}		
\newcommand{\apjs}{ApJS}		
\newcommand{\na}{New Astronomy} 
\newcommand{\mnras}{MNRAS}		

\newcommand{\pasj}{PASJ}		

\begin{document}


\title[Tidal evolution of halos]{Tidal evolution of cored and cuspy dark matter halos}

\author{Xiaolong Du\,\orcidlink{0000-0003-0728-2533}}
\email{xdu@astro.ucla.edu}
\affiliation{Department of Physics and Astronomy, University of California, Los Angeles, CA 90095, USA}
\affiliation{Carnegie Observatories, 813 Santa Barbara Street, Pasadena, CA 91101, USA}

\author{Andrew Benson\,\orcidlink{0000-0001-5501-6008}}
\email{abenson@carnegiescience.edu}
\affiliation{Carnegie Observatories, 813 Santa Barbara Street, Pasadena, CA 91101, USA}

\author{Zhichao Carton Zeng\,\orcidlink{0000-0002-6008-5144}}
\affiliation{Department of Physics, The Ohio State University, 191 W. Woodruff Ave., Columbus, OH 43210, USA}
\affiliation{Center for Cosmology and Astroparticle Physics, The Ohio State University, 191 W. Woodruff Ave., Columbus, OH 43210, USA}

\author{Tommaso Treu\,\orcidlink{0000-0002-8460-0390}}
\affiliation{Department of Physics and Astronomy, University of California, Los Angeles, CA 90095, USA}

\author{Annika H. G. Peter\,\orcidlink{0000-0002-8040-6785}}
\affiliation{Department of Physics, The Ohio State University, 191 W. Woodruff Ave., Columbus, OH 43210, USA}
\affiliation{Center for Cosmology and Astroparticle Physics, The Ohio State University, 191 W. Woodruff Ave., Columbus, OH 43210, USA}
\affiliation{Department of Astronomy, The Ohio State University, 140 W. 18th Ave., Columbus, OH 43210, USA}
\affiliation{School of Natural Sciences, Institute for Advanced Study, 1 Einstein Drive, Princeton, NJ 08540}

\author{Charlie Mace\,\orcidlink{0000-0002-9419-6547}}
\affiliation{Department of Physics, The Ohio State University, 191 W. Woodruff Ave., Columbus, OH 43210, USA}
\affiliation{Center for Cosmology and Astroparticle Physics, The Ohio State University, 191 W. Woodruff Ave., Columbus, OH 43210, USA}

\author{Fangzhou Jiang\,\orcidlink{0000-0001-6115-0633}}
\affiliation{Kavli Institute for Astronomy and Astrophysics, Peking University, Beijing 100871, People’s Republic of China}
\affiliation{Carnegie Observatories, 813 Santa Barbara Street, Pasadena, CA 91101, USA}
\affiliation{TAPIR, California Institute of Technology, Pasadena, CA 91125, USA}

\author{Shengqi Yang\,\orcidlink{0000-0002-0782-9116}}
\affiliation{Carnegie Observatories, 813 Santa Barbara Street, Pasadena, CA 91101, USA}

\author{Charles Gannon\,\orcidlink{0009-0009-0443-3181}}
\affiliation{University of California, Merced, 5200 N Lake Road, Merced, CA 95341, USA}

\author{Daniel Gilman\,\orcidlink{0000-0002-5116-7287}}
\affiliation{Department of Astronomy $\&$ Astrophysics, University of Chicago, Chicago, IL 60637, USA}

\author{Anna. M. Nierenberg\,\orcidlink{0000-0001-6809-2536}}
\affiliation{University of California, Merced, 5200 N Lake Road, Merced, CA 95341, USA}

\author{Ethan O.~Nadler\,\orcidlink{0000-0002-1182-3825}}
\affiliation{Carnegie Observatories, 813 Santa Barbara Street, Pasadena, CA 91101, USA}
\affiliation{Department of Physics $\&$ Astronomy, University of Southern California, Los Angeles, CA 90007, USA}

\begin{abstract}
The internal structure and abundance of dark matter halos and subhalos are powerful probes of the nature of dark matter. In order to compare observations with dark matter models, accurate theoretical predictions of these quantities are needed. We present a fast and accurate method to describe the tidal evolution of subhalos within their parent halo, based on a semi-analytic approach. We first consider idealized $N$-body simulations of subhalos within their host halo, using a generalized mass density profile that describes their properties in a variety of dark matter models at infall, including popular warm, cold, and self-interacting ones. Using these simulations we construct tidal ``tracks" for the evolution of subhalos based on their conditions at infall. Second, we use the results of these simulations to build semi-analytic models for tidal effects, including stripping and heating and implement them within the code {\sc galacticus}. Our semi-analytic models can accurately predict the tidal evolution of both cored and cuspy subhalos, including the bound mass and density profiles, providing a powerful and efficient tool for studying the post-infall properties of subhalos in different dark matter models.
\end{abstract}

\maketitle


\section{Introduction}\label{sec:intro}

The vast majority of the matter density of the Universe ($\sim 85\%$) is known to be non-baryonic, i.e.\ made of something other than the quarks and baryons of the standard model of particle physics. Understanding the fundamental physical nature of dark matter (DM) has been a long standing goal of physics and cosmology. The commonly accepted model postulates that DM is composed by a massive non-relativistic particle (sometimes known as the Weakly Interacting Massive Particle, or WIMP), and behaves cosmologically as Cold Dark Matter (CDM).

However, while the CDM model has shown excellent agreement with observations on large scales, such as the Cosmic Microwave Background and large-scale structure of galaxy distributions, reproducing some observations on subgalactic scales is challenging within the model~\cite{BBK:2017,Buckley:2017ijx}. Examples of these challenges include the cusp-core problem~\cite{Moore:1994yx,deBlok:2001hbg,Newman:2013}, the too-big-to fail problem ~\cite{MBK:2011}, and diversity problem~\cite{Oman:2015xda,Read:2018fxs}. To solve these issues, a number of alternative DM models have been proposed, including, e.g. warm dark matter~\cite{Dodelson:1994,Shi:1999,Abazajian:2001,Dolgov:2000ew}, fuzzy dark matter (FDM)~\cite{Sin:1994,Hu:2000ke,Schive:2014dra,Schive:2014hza,Marsh:2015,Hui:2016ltb}, self-interacting dark matter (SIDM)~\cite{Spergel:1999mh,Tulin:2017ara}, and primordial black holes~\cite{Chapline:1975,Carr:2020xqk,Green:2020jor}. 

A particularly powerful probe of the nature of DM is the abundance and internal structure of halos and subhalos. These are the bound hierarchical structures that form as a result of gravity, as the Universe evolves. Alternate dark matter models predict different density profiles and abundance for small DM halos and subhalos. The demographics and internal structure of halos and subhalos already provide some of the most stringent limits on alternative dark matter models, including from observations of Milky Way satellites~\cite{Jethwa:2016gra,Kim:2017iwr,Hayashi:2020syu,Newton:2020cog,Nadler:2021,Nadler:2021dft,Dalal:2022rmp,Dekker:2021scf,Kim:2021zzw,Esteban:2023xpk,Akita:2023yga} and strong gravitational lensing \cite{Gilman:2019nap,Enzi:2020ieg,Nadler:2021dft,Zelko:2022,Dike:2022heo,Powell:2023jns,Vegetti:2023mgp}.

In order to interpret the observation of DM halos and subhalos in terms of reliable constraints on the nature of DM, it is essential to have accurate theoretical predictions to compare with. Numerical simulations have been shown to be an essential tool to connect the fundamental physics to the growth of structure in the Universe. With the rapid development in the computing resources and techniques, state-of-the-art cosmological simulations are pushing the boundary of numerical capability, i.e.\ to resolve the smallest dark matter halos~\cite{Diemand:2008in,Springel:2008cc,Delos:2022yhn,Garrison-Kimmel:2013eoa,Griffen:2015rqk,Wang:2019ftp} and to simulate a large cosmic volume~\cite{Riebe:2011,Schaye:2023jqv}. 

In practice, numerical simulations are always limited by resolution, especially when we focus on the evolution of subhalos. Unlike isolated halos, subhalos are subject to frequent tidal interactions with their host halos, which leads to mass loss, tidal heating, and possible disruption. Even state-of-the-art simulations may suffer from artificial disruption of subhalos~\cite{vandenBosch:2018tyt}, which may lead to biased results when comparing with observations. 

One way to control numerical artifacts is to run idealized simulations, in which case only one subhalo is evolved in the host potential so that one can run high-resolution simulations with manageable computing resource. Using such methods, van den Bosch and Ogiya~\cite{vandenBosch:2018tyt} carried out a detailed study of numerical artifacts in simulations resulting from gravitational softening, discreteness noise, and two-body relaxation. They also derived the requirements for obtaining properly converged results. Using a similar approach, Aguirre-Santaella et al.~\cite{Aguirre-Santaella:2022kkm} studied the evolution of subhalos in a realistic Milky Way-like host potential including contributions from the galactic bulge and disk. However, in order to compute statistical properties of subhalos, one needs a model to describe the tidal evolution of subhalos so that one could look at the evolution of a population of subhalos within the context of a cosmological model. In order to achieve this, one can either build analytic models using appropriate approximations, or train non-parametric models such as the one presented in Ref.~\cite{Ogiya:2019del}.  

In this work, we first run idealized $N$-body simulations to study the evolution of DM halos with different initial density profiles in a host gravitational potential, aiming to include a broad range of dark matter profiles that may be produced by different DM models. Previous studies~\cite{Hayashi:2002qv,Penarrubia:2010jk,Green:2019zkz,Errani:2020wgn} have shown that, as subhalos evolve in the host, their maximum circular velocity, $V_{\mathrm{max}}$,~\footnote{The circular velocity is defined as $V=\sqrt{G M(r)/ r}$.} and the radius at which this maximum is reached, $R_{\mathrm{max}}$, follow a universal ``tidal track" for a specific initial density profile. Both $V_{\rm max}/V_{\rm max,0}$ and $R_{\rm max}/R_{\rm max,0}$ are functions of fractional mass remaining in the subhalo and are not sensitive to \emph{how} mass is stripped from the subhalos. We calculate the tidal tracks for different initial dark matter profiles, including DM halos with cored profiles and those with extremely cuspy profiles. Cored profiled can result from non-gravitational interactions between DM particles, e.g.\ in FDM~\cite{Schive:2014dra,Schive:2014hza} and SIDM~\cite{Spergel:1999mh,Tulin:2017ara} models, while extremely cuspy profiles are found in SIDM model when a halo undergoes core collapse~\cite{Balberg:2002ue,Ahn:2004xt,Koda:2011yb,Yang:2022zkd,Outmezguine:2022bhq}. Notably, Penarrubia {\it et al.}~\cite{Penarrubia:2010jk} has found that for DM profiles with different inner slopes, the tidal track is significantly different. We verify such dependence on the inner slope and also investigate the influence of the density slope at larger radii.

We then build improved semi-analytic models (SAMs) for the tidal effects, including tidal stripping and tidal heating, and calibrate these models to the idealized simulations. In previous work~\cite{Taylor:2000zs,Taffoni:2003sr,Pullen:2014gna,Yang:2022zkd}, similar SAMs have been used to describe the tidal evolution of DM halos initialized with Navarro-Frenk-White (NFW) density profiles~\cite{1997ApJ...490..493N}. To model the density evolution of a subhalo due to tidal heating, a commonly used approach is to estimate the heating energy rate using the impulse approximation~\cite{Gnedin:1997vp}. See also Ref.~\cite{Stucker:2022fbn} for another approach using the adiabatic limit. From the heating energy, the expansion of mass shells in a subhalo can be solved and converted to the change in density profile \cite{Taylor:2000zs,Pullen:2014gna}. In one of our previous studies ~\cite{2022MNRAS.517.1398B}, we showed that by including the contribution from the second-order perturbation in the heating energy, we can reproduce the tidal track for NFW halos accurately. In this work we will extend these models to other dark matter profiles.

This paper is organized as follows. In Sec.~\ref{sec:setup}, we describe the setup of our idealized simulations. In Sec.~\ref{sec:results}, we show our results from the simulations and give fitting functions for the tidal tracks and density transfer functions assuming different initial density profiles. In Sec.~\ref{sec:sams}, we describe our semi-analytic models and show the calibrations of the model parameters. Conclusions and discussions are given in Sec.~\ref{sec:cons}.

\section{Simulation Setup}\label{sec:setup}

We simulate the evolution of a subhalo in a static host halo potential. For the definition of virial radius, $r_\mathrm{vir}$, we make use of the spherical collapse model~\cite{Percival:2005vm}
\begin{equation}
\overline{\rho}(<r_{\mathrm{vir}})=\Delta_{\mathrm{vir}}\,\rho_m = \Delta_{\mathrm{vir}}\,\Omega_m\,\frac{3 H_0^2}{8 \pi G}.
\label{eq:def_vir}
\end{equation}
At $z=0$, $\Delta_{\mathrm{vir}}=329.621$. Cosmological parameters from \cite{Planck:2018vyg} are adopted, i.e. $\Omega_m=0.3153$, $\Omega_{\Lambda}=0.68470$, $H_0=67.36\,{\mathrm{km/s/Mpc}}$.

\subsection{Initial conditions}\label{subsec:ICs}

The host is defined to have an NFW density profile~\cite{1997ApJ...490..493N}
\begin{equation}
\rho_{\mathrm NFW}(r) = \frac{\rho_0}{\frac{r}{r_\mathrm{s}}\left(1+\frac{r}{r_\mathrm{s}}\right)^2},
\label{eq:NFW}
\end{equation}
where $\rho_0$ is the characteristic density and $r_\mathrm{s}$ is the scale radius. The acceleration due to the host is then given by
\begin{eqnarray}
\mathbf{a}_{\mathrm NFW}(r) = -4\pi\rho_0 r_\mathrm{s}^3\left[\ln\left(1+\frac{r}{r_\mathrm{s}}\right)-\frac{\frac{r}{r_\mathrm{s}}}{1+\frac{r}{r_\mathrm{s}}}\right] \nonumber \\
\frac{1}{r^2+r_{\mathrm{softening}}^2}\frac{\mathbf{r}}{r},
\label{eq:NFW_acc}
\end{eqnarray}
where the term involving $r_{\mathrm{softening}}$, the gravitational softening length, is introduced to soften the potential in the inner regions~\cite{Barnes:2012qg}. For consistency, the value of $r_{\mathrm{softening}}$ is taken to be the same as that used for gravitational softening between particles. For the host, we choose
\begin{eqnarray}
\rho_0           &=& 3.797 \times 10^6\,\mathrm{M}_{\odot}/{\mathrm{kpc}}^3,\\
r_\mathrm{s}     &=& 23.69\,\mathrm{kpc},\\
r_{\mathrm{vir}} &=& 263.2\,\mathrm{kpc},\\
M_{\mathrm{vir}} &=& 10^{12}\,\mathrm{M}_{\odot}.
\label{eq:NFW_par}
\end{eqnarray}
The host mass and concentration, $c_{\mathrm{host}}=r_{\mathrm{vir}}/r_{\mathrm{s}}$, are chosen to represent the typical values for the Milky Way DM halo (see e.g.~\cite{McMillan:2011wd}).

At the initial time the subhalo is assumed to be spherically symmetric and has a density profile described by~\cite{Hernquist:1990be,Zhao:1995cp,Kravtsov:1997dp}:
\begin{equation}
\rho(r)=\frac{\rho_0}{\left(\frac{r}{r_\mathrm{s}}\right)^{\gamma}\left[1+\left(\frac{r}{r_\mathrm{s}}\right)^{\alpha}\right]^{(\beta-\gamma)/\alpha}}.
\label{eq:prof_abg}
\end{equation}
The above parametrization with three free parameters are also known as the ``Nuker" model which was first used to describe the surface brightness profile of galactic nuclei~\cite{1995AJ....110.2622L}. To enforce that the subhalo has a finite total mass, we truncate the subhalo density profile at $r>r_{\mathrm{vir}}$ following \cite{Kazantzidis:2003im}
\begin{eqnarray}
\rho(r)=\frac{\rho_0}{c^{\gamma}(1+c^{\alpha})^{(\beta-\gamma)/\alpha}}\left(\frac{r}{r_{\mathrm{vir}}}\right)^{\kappa}\exp\left(-\frac{r-r_{\mathrm{vir}}}{r_{\mathrm{decay}}}\right),
\label{eq:trunc}
\end{eqnarray}
where $c=r_{\mathrm{vir}}/r_\mathrm{s}$ is the halo concentration and
\begin{equation}
\kappa=-\frac{\gamma+\beta c^{
\alpha}}{1+c^{\alpha}}+\frac{r_{\mathrm{vir}}}{r_{\mathrm{decay}}}.
\label{eq:kappa}
\end{equation}
We take $r_{\mathrm{decay}}/r_{\mathrm{vir}}=0.1$, with which the total mass of subhalo is $1.02-1.2 M_{\mathrm{vir,sub}}$ depending on the parameters $(\alpha,\beta,\gamma)$. Other parameters in the subhalo density profile are taken to be
\begin{eqnarray}
r_\mathrm{s,sub}   &=&1.279\,\mathrm{kpc},
\label{eq:rs}\\
r_{\mathrm{vir,sub}}&=&26.32\,\mathrm{kpc},
\label{eq:rvir}\\
M_{\mathrm{vir,sub}}&=&10^{9}\,\mathrm{M}_{\odot}.
\label{eq:Mvir}
\end{eqnarray}
The above subhalo mass and concentration are typical values for dwarf satellite galaxies in the Milky Way (see e.g.~\cite{Read:2019}). For this choice of subhalo mass, the dynamical friction effect is not important~(see Sec.~\ref{subsec:orbit}), thus it is appropriate to simulate its tidal evolution assuming a static host potential. Note that we fix the virial mass of the subhalo, such that $\rho_0$ will differ for different combinations of $(\alpha,\beta,\gamma)$.

Given the density profile of a subhalo, an $N$-body realization is generated by sampling particle positions from that density profile. For the initial velocities of particles we assume an isotropic velocity dispersion and use Eddington's formula \cite{Eddington:1916} to compute the velocity distribution function as a function of radius, taking into account the effects of gravitational softening using the approach of \cite{Barnes:2012qg}. Velocities are then sampled from that distribution function at the position of each particle. We take a particle mass of $M_\mathrm{p}=10^2\, \mathrm{M}_{\odot}$ so that the subhalo contains $N\sim 10^7$ particles within its virial radius initially. We limit our analysis to the simulation outputs when the subhalo is still resolved by at least $10^4$ particles so that its density profile can be well measured. For the choice of softening length, we follow Ref.~\cite{Ogiya:2019del}, in which it is suggested that for $N=10^6$ a value of $r_{\mathrm{softening}}= 0.0003 r_{\mathrm{vir,sub}}$ is sufficient to mitigate the numerical artifacts~\cite{vandenBosch:2018tyt}. We rescale the softening length according to the particle number, i.e.\ $r_{\mathrm{softening}}\propto N_{p}^{1/3}$~\cite{vanKampen:2000px} and take
\begin{equation}
r_{\mathrm{softening}}=1.39 \times 10^{-4} r_{\mathrm{vir,sub}}=0.00367\,{\mathrm{kpc}}.
\label{eq:r_soft}
\end{equation}
As we will discuss below, for the case $\gamma=1.5$, we also check the convergence of our results by using different numbers of particles and varying the softening length.

The subhalo is initially placed at the apocenter of its orbit at $R=0.7 r_{\mathrm{vir,host}}$ as in Ref.~\cite{Penarrubia:2010jk} and given a tangential velocity $v_\mathrm{t}$, where $v_\mathrm{t}$ is determined by assuming different pericentric/apocentric ratios, $R_\mathrm{p}/R_\mathrm{a}$. For each combination, we run the simulation with different values of $R_\mathrm{p}/R_\mathrm{a}$
so that the tidal tracks are well measured at early times and at the same time include the regime where the subhalos have been heavily stripped.

\subsection{Orbital and tidal evolution}\label{subsec:orbit_evo}

The {\sc Gadget-4} code ~\cite{Springel:2020plp} is used to simulate the evolution of the subhalo within the static host potential. Instead of the TreePM method commonly used in previous simulations, we make use of a new feature in {\sc Gadget-4}, the so-called fast multipole method (FMM) which has the advantage that the momentum is better conserved. We did not switch on the ``PMGRID'' option since we find for the simulations we perform the pure FMM method is faster and more accurate. The parameters controlling the accuracy of force computation and time integration are taken to be
\begin{center}
\begin{tabular}{rcl}
\textsc{ErrTolIntAccuracy}     &=&0.01,\\
\textsc{MaxSizeTimestep}       &=&0.01,\\
\textsc{MinSizeTimestep}       &=&0.00,\\
\textsc{TypeOfOpeningCriterion}&=&1,   \\
\textsc{ErrTolTheta}           &=&0.1, \\
\textsc{ErrTolThetaMax}        &=&1.0, \\
\textsc{ErrTolForceAcc}        &=&0.001.
\label{eq:GADGET_par}
\end{tabular}
\end{center}
Our tests show that the above values are sufficient for most of the cases we have run (but see Sec.~\ref{subsec:convergence} for the exceptional case with an extremely cuspy profile). Details of each parameter can be found in the {\sc Gadget-4} documentation.~\footnote{\url{https://wwwmpa.mpa-garching.mpg.de/gadget4/gadget4_manual.pdf}}

The particles are evolved using the hierarchical time integration scheme introduced in {\sc Gadget-4}. The time step, $\Delta t_\mathrm{i}$ for particle i is determined based on its acceleration $\mathbf{a}_\mathrm{i}$:
\begin{equation}
\Delta t_\mathrm{i} = \min \left\{\sqrt{\frac{2\eta r_{\rm softening}}{|\mathbf{a_\mathrm{i}}|}},\Delta t_{\rm max}\right\},
\label{eq:time_step}
\end{equation}
where $\eta$ is set by the parameter ``\textsc{ErrTolIntAccuracy},'' and $\Delta t_{\rm max}$ is set by ``\textsc{MaxSizeTimestep}.'' For the extremely cuspy case with $\gamma=1.5$, we find that the results for the tidal track are not converged with the fiducial settings and an extremely small $\eta$ is needed to achieve convergence, so we also test a fixed, global time step in which case all particles are forced to have the same time step size determined by the minimum over $\Delta t_\mathrm{i}$ (see Sec.~\ref{subsec:convergence}).

The subhalo is evolved for $37\,{\mathrm{Gyr}}$. The particle data are output every $0.09778\,{\mathrm{Gyr}}$. For each snapshot, we performed a self-binding analysis (see Sec.~\ref{sec:results}) to identify particles that remain bound to the subhalo, and computed the rotation curve of the subhalo using those bound particles.

\subsection{Analysis}\label{subsec:analysis}

A reliable self-binding analysis is very important for identifying subhalos and determining their properties (such as the bound mass). Typically, in cosmological simulations, various halo finders such as Rockstar \cite{2013ApJ...762..109B} and AHF \cite{2009ApJS..182..608K} have been employed in both cosmological and idealized simulations 
(see, for example \cite{2022MNRAS.513.4845Z}).
Robust detection and characterization of subhalos in such simulations is challenging---for example, recent studies show that the commonly used Rockstar halo finder fails to find a non-negligible fraction of subhalos \cite{2023arXiv230500993D,2023arXiv230810926M}. In the present work we have the advantage that \emph{all} particles begin as members of the subhalo---we therefore know the starting point which allows us to proceed in a more careful and controlled manner from one snapshot to the next. Our self-binding algorithm is implemented in the {\sc Galacticus} code \cite{2012NewA...17..175B} to perform this analysis.

For each snapshot of a simulation, we carry out a self-binding analysis to determine which particles remain bound to the subhalo. This is an iterative process, which proceeds as follows:
\begin{enumerate}
    \item At the beginning of simulation, all particles are assumed to be gravitationally bound to the subhalo. For snapshots at later times, we use the bound/unbound status of particles from the previous snapshot in the first iteration of our algorithm.
    \item The center-of-mass position ${\mathbf r}_{\rm cm}$ and velocity $\mathbf{v}_{\rm cm}$ are determined from all bound particles.
    \item The gravitational potential energy and kinetic energy are computed for each bound particle:
    \begin{eqnarray}
    E_\mathrm{p,i} &=& -\sum_{\mathrm{i} \neq \mathrm{j}} m_\mathrm{p} \Phi_\mathrm{ij}, \label{eq:E_p}\\
    E_\mathrm{k,i} &=& \frac{1}{2} m_\mathrm{p} (\mathbf{v}_\mathrm{i}-\mathbf{v}_{\rm cm})^2.
    \label{eq:E_k}
    \end{eqnarray}
    Here $\Phi_\mathrm{ij}$ is the gravitational potential between particles i and j taking into account the effect of the gravitational softening used in the simulation. Note that when computing $E_\mathrm{p,i}$, only contributions from particles identified as being bound to the subhalo in the previous iteration are included.
    \item Particles with positive total energy, $E_\mathrm{t,i}=E_\mathrm{p,i}+E_\mathrm{k,i}$ are considered unbound and excluded from later analysis.
    \item Repeat steps (2)--(4) until the following criterion is satisfied:
    \begin{equation}
    \frac{\left|M_{\rm bound}-M_{\rm bound}^{\rm previous}\right|}{\left(M_{\rm bound}+M_{\rm bound}^{\rm previous}\right)/2} < \epsilon.
    \label{eq:stop_criteria}
    \end{equation}
    For all analyses, we take $\epsilon=10^{-3}$.
    \item The center of the subhalo is determined by searching for the particle with the most negative $E_\mathrm{p,i}$, which corresponds to the position with the highest density. Similarly, the representative velocity of the subhalo is determined by the particle with the most negative $E_\mathrm{v,i}$. Here $E_\mathrm{v,i}$ is computed in the same way as $E_\mathrm{p,i}$ but replacing the coordinates of particles with velocities, following the approach of \cite{2005MNRAS.358..551B}.
\end{enumerate}
After completing the analysis described above, the center of the subhalo in phase space is then used to compute the density profile and velocity dispersion profile of the subhalo as a function of radius.

This procedure is similar to the one used by van den Bosch and Ogiya~\cite{vandenBosch:2018tyt}. In Ref.~\cite{vandenBosch:2018tyt}, the authors use the $5\%$ most-bound particles to determine the center-of-mass properties of the subhalo. In our analysis, we compute ${\mathbf r}_{\rm cm}$ and ${\mathbf v}_{\rm cm}$ using all bound particles, which in general gives smoother results for $M_{\rm bound}$. Furthermore, our convergence criteria differ slightly from that of Ref.~\cite{vandenBosch:2018tyt}, based on the bound mass rather than the center-of-mass position and velocity. 

The evolution of a subhalo's density profile as it orbits within a host potential can be characterized by its $V_{\mathrm{max}}$ and $R_{\mathrm{max}}$.
As the subhalo evolves, it is subjected to the tidal force of the host halo. $V_{\rm max}/V_{\rm max,0}$ and $R_{\rm max}/R_{\rm max,0}$ evolve along a so-called called tidal track which is largely independent of precisely how mass is stripped away from the subhalos \citep{Hayashi:2002qv,Penarrubia:2010jk,Green:2019zkz,Errani:2020wgn}. Here $V_{\rm max,0}$ and $R_{\rm max,0}$ are the initial values of these parameters. While the tidal track is insensitive to the host potential, it does show significant differences for different initial density profiles~\cite{Penarrubia:2010jk}. Therefore, finding the tidal tracks for different dark matter profiles is crucial for modeling how subhalo evolution in a host potential may differ under various assumptions for the nature of the dark matter particle.

We compute $V_{\mathrm{max}}$ and $R_{\mathrm{max}}$ from bound particles for each snapshot of our simulations. In practice, after finding the center of the subhalo, we compute the mass profile using radial bins. At small radii, i.e. $r < 4 r_{\rm kernel}$ with $r_{\rm kernel}=2.8 r_{\rm sofening}$, linear bins are used to reduce the discreteness noise. The bin width is taken to be $r_{\rm kernel}$. At larger radii, logarithmic bins are used and the bin width is taken to be $\Delta \log_{10} R = 0.1$. The mass profile data is then super-sampled by a factor of $40$ using a cubic spline interpolation to determine $V_{\mathrm{max}}$ and $R_{\mathrm{max}}$. We find that this approach allows us to obtain more accurate $V_{\mathrm{max}}$ and $R_{\mathrm{max}}$ that are less affected by discreteness noise.

\section{Results}\label{sec:results}

\subsection{Tidal tracks}\label{subsec:track}

\begin{figure*}
\includegraphics[width=\columnwidth]{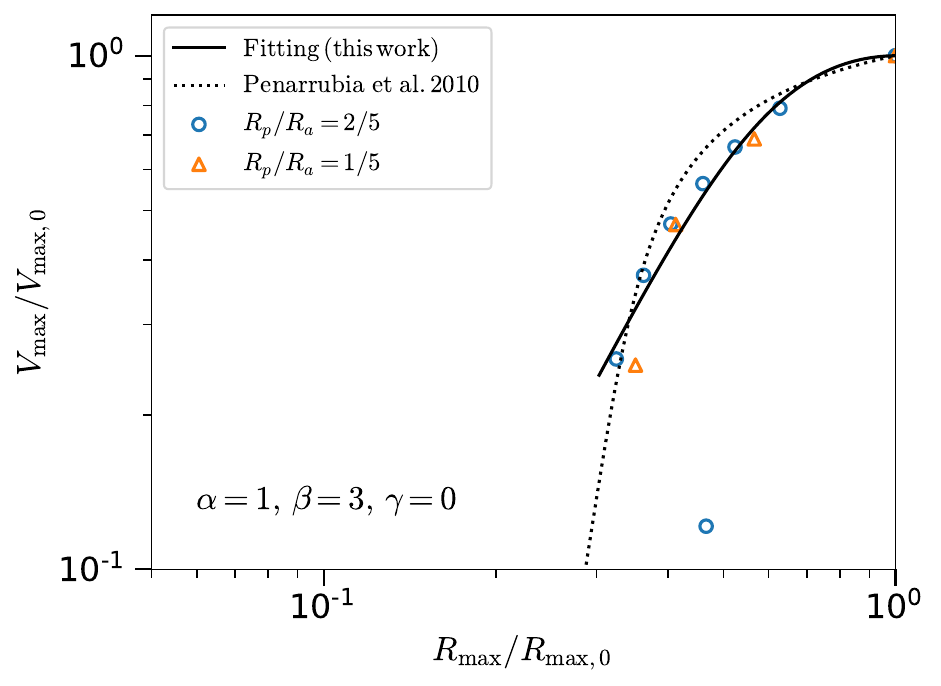}
\includegraphics[width=\columnwidth]{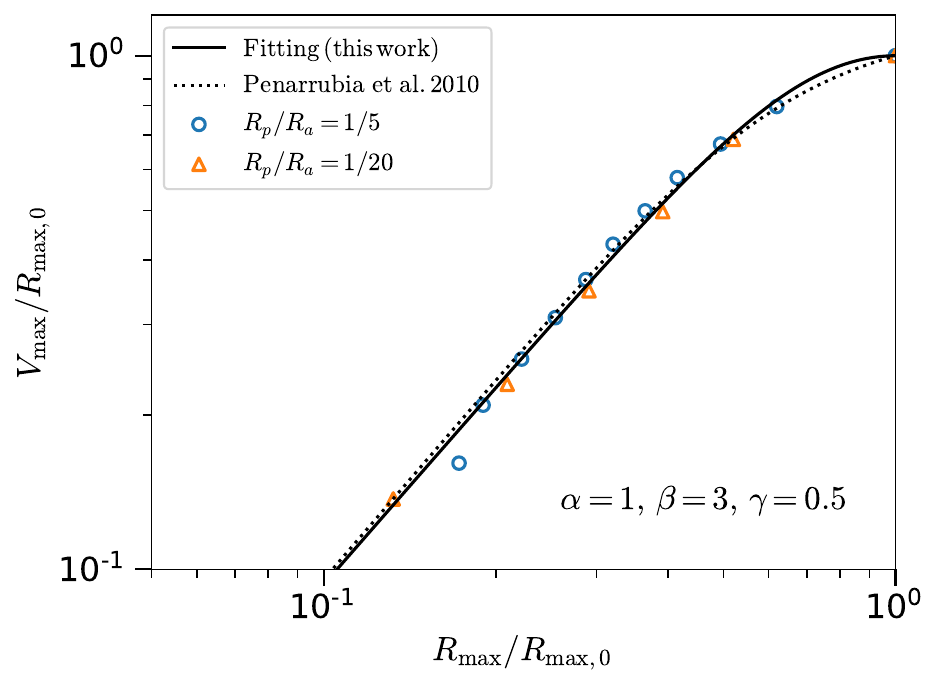}
\includegraphics[width=\columnwidth]{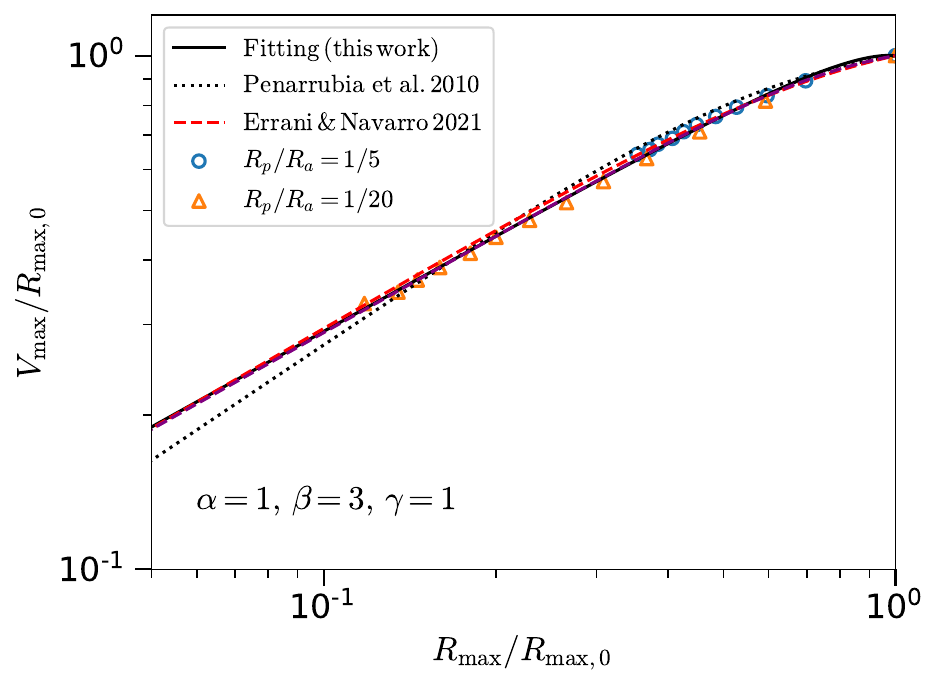}
\includegraphics[width=\columnwidth]{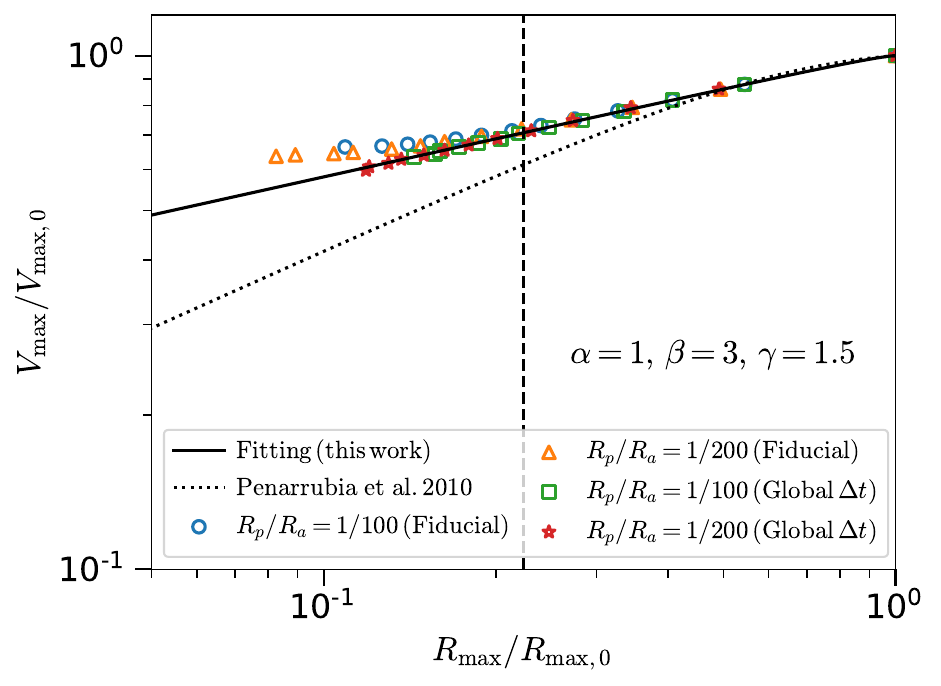}
\caption{Tidal tracks for $(\alpha,\beta)=(1,3)$ and different values of $\gamma$ and pericentric/apocentric ratios, $R_p/R_a$, from $N$-body simulations using {\sc Gadget-4}.
Lines show the fitting functions found of \protect\cite{Penarrubia:2010jk} (dotted), \protect\cite{Errani:2020wgn} (red dashed, for $\gamma=1$ only), and this work (solid). Note that for the case with $\gamma=0.5$, the last data point, i.e.\ the one with the lowest $V_{\mathrm{max}}/V_{\mathrm{max},0}$ has been excluded from fitting. For $\gamma=1.5$ (lower right panel), the results from the fiducial runs (circles and triangles) are not converged below $R_{\mathrm{max}}/R_{\mathrm{max},0}\sim 0.22$ (vertical dashed line). Using a global time step size (squares and stars) leads to better converged results, see the discussions in the Sec.~\ref{subsec:convergence}.}
\label{fig:tidal_track_GADGET}
\end{figure*}

\begin{figure*}
\includegraphics[width=0.95\columnwidth]{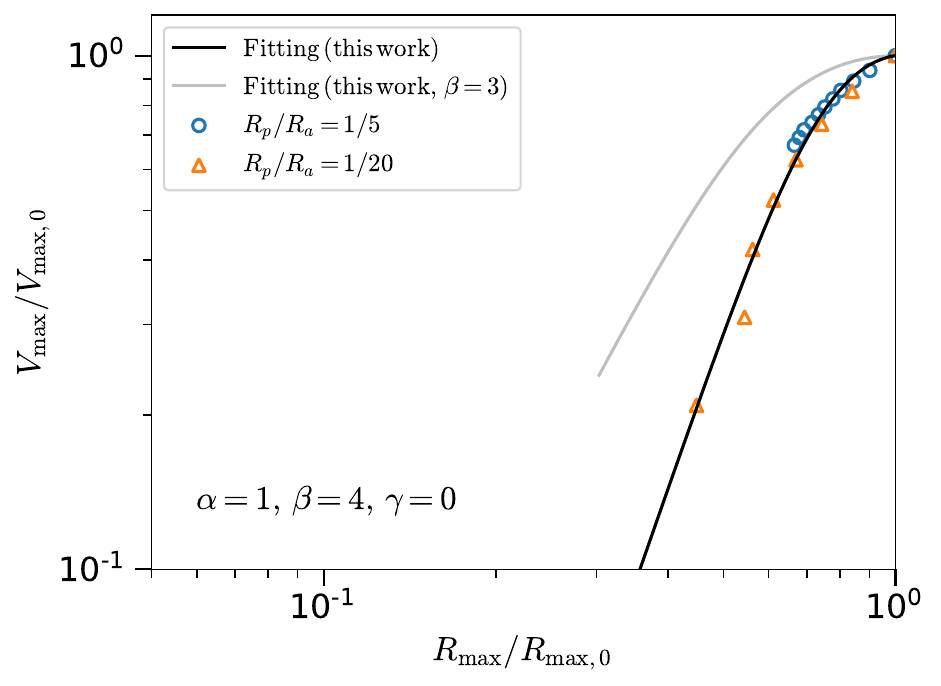}
\includegraphics[width=0.95\columnwidth]{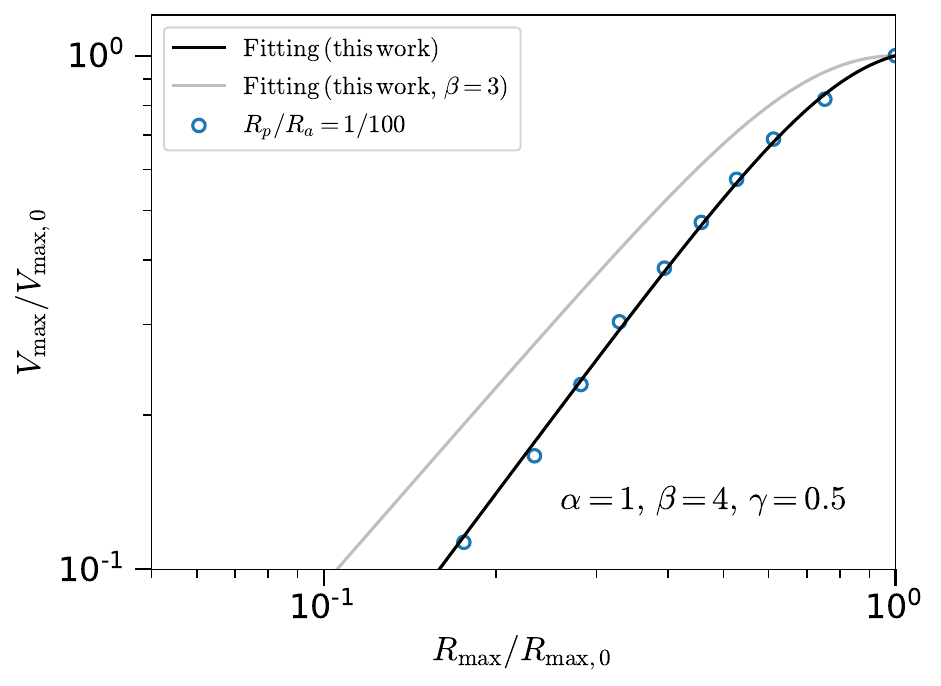}
\includegraphics[width=0.95\columnwidth]{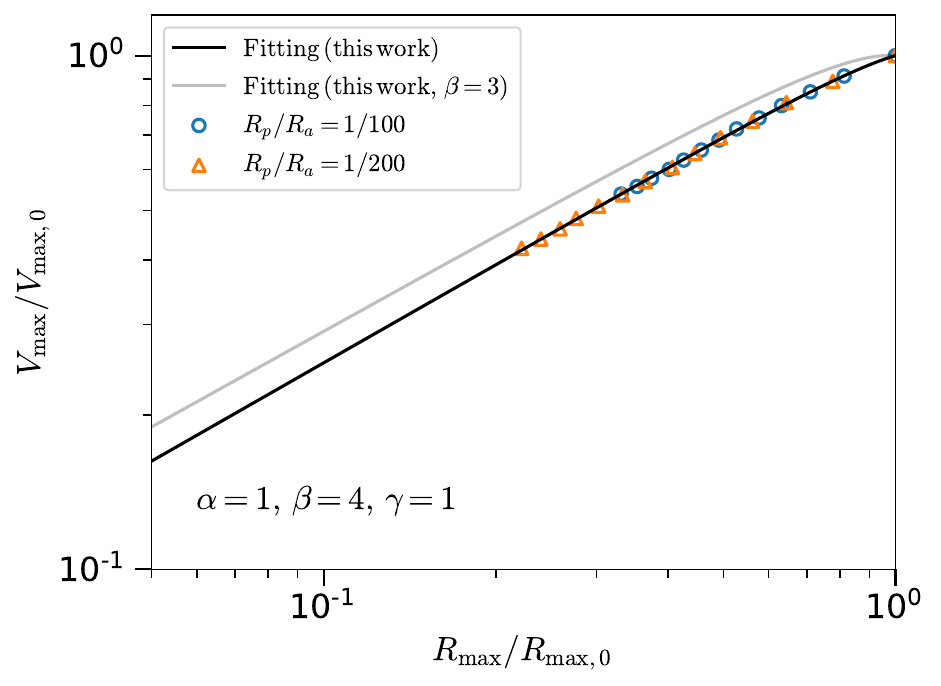}
\includegraphics[width=0.95\columnwidth]{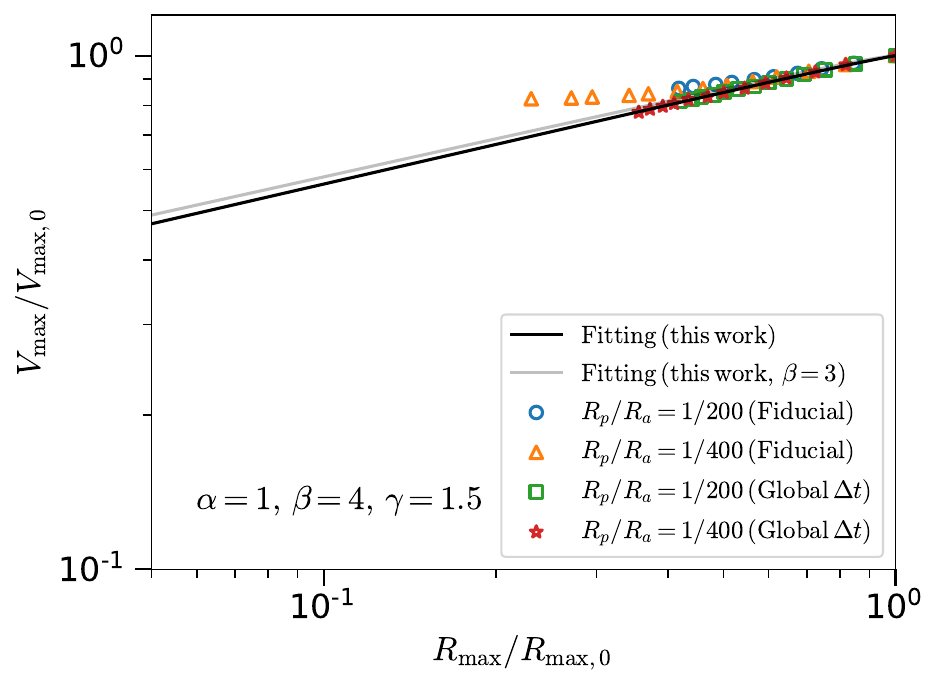}
\caption{Tidal tracks for $(\alpha,\beta)=(1,4)$ and different values of $\gamma$ from $N$-body simulations using {\sc Gadget-4}. Solid black lines show the fitting functions found in this work. For comparisons, the fitting functions for the same $(\alpha, \gamma)$, but $\beta=3$ are also shown in each panel (light gray lines).}
\label{fig:tidal_track_GADGET2}
\end{figure*}

\begin{figure*}
\includegraphics[width=0.95\columnwidth]{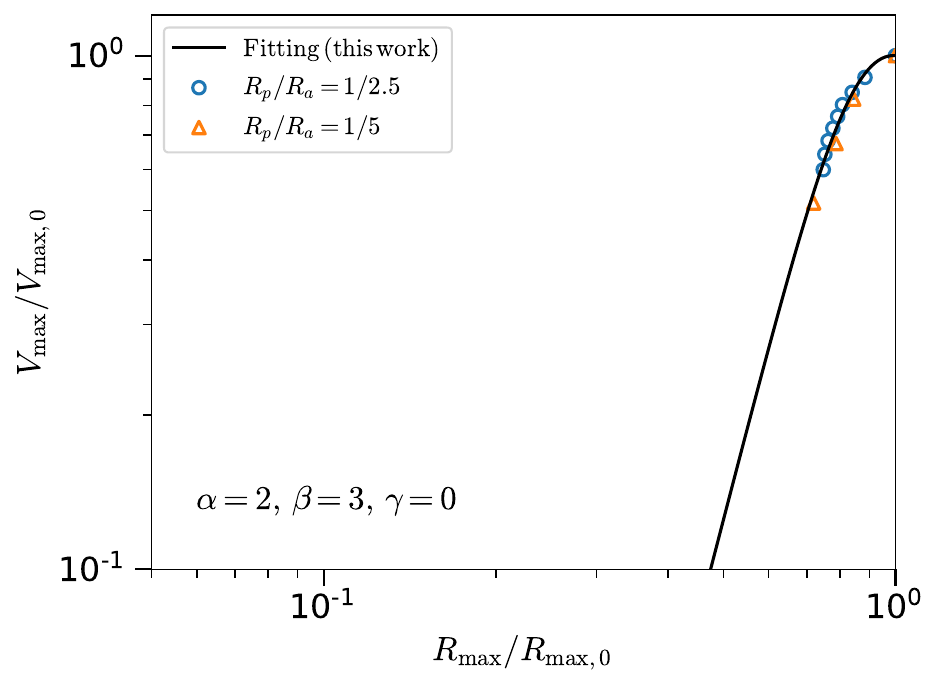}
\includegraphics[width=0.95\columnwidth]{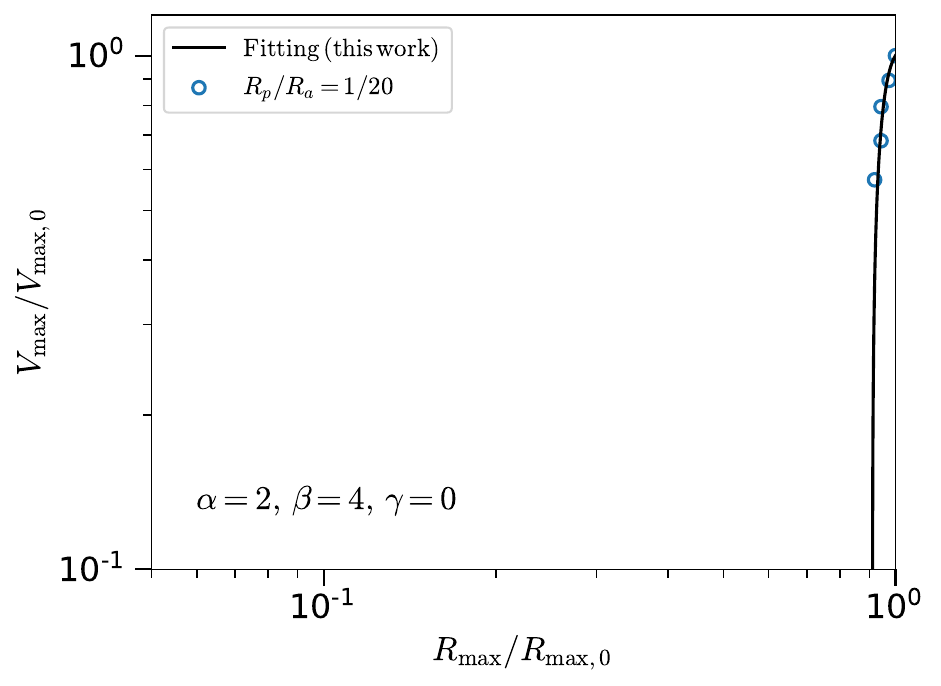}
\caption{Tidal tracks for $(\alpha,\gamma)=(2,0)$ and different values of $\beta$ from $N$-body simulations using {\sc Gadget-4}. Solid lines show the fitting functions found in this work.}
\label{fig:tidal_track_GADGET3}
\end{figure*}

The tidal tracks, $V_{\mathrm{max}}/V_{\mathrm{max},0}$ versus $R_{\mathrm{max}}/R_{\mathrm{max},0}$, for different combinations of $(\alpha,\beta,\gamma)$ are shown in Figs.~\ref{fig:tidal_track_GADGET}, \ref{fig:tidal_track_GADGET2} and \ref{fig:tidal_track_GADGET3}. First, we keep $\alpha$ and $\beta$ fixed at $1$ and $3$ (the values for an NFW profile), respectively, and change $\gamma$ (the logarithmic slope of density profile at small radii), see Fig.~\ref{fig:tidal_track_GADGET}. As can be seen, for $\gamma \leq 1$, the results from {\sc Gadget-4} are in good agreement with the fitting functions found by \cite{Penarrubia:2010jk} (dotted) and \cite{Errani:2020wgn} (red dashed, for $\gamma=1$ only). However, for $\gamma=1.5$ we find that the tidal track obtained from {\sc Gadget-4} simulations deviates from the \cite{Penarrubia:2010jk} fitting function at late times (i.e. for small values of $V_{\mathrm{max}}/V_{\mathrm{max},0}$ and $R_{\mathrm{max}}/R_{\mathrm{max},0}$). Increasing or decreasing the gravitational softening length by a factor of $3$ slightly changes our results, but can not explain the large deviation from the fitting curve.

For $\gamma=0$, we find that as the subhalo evolves in the host potential, both $V_{\mathrm{max}}/V_{\mathrm{max},0}$ and $R_{\mathrm{max}}/R_{\mathrm{max},0}$ decrease (from the upper right corner toward the lower left corner). However, as tidal evolution proceeds further, there exists a turnaround at $R_{\mathrm{max}}/R_{\mathrm{max},0}\sim 0.3$, after which $R_{\mathrm{max}}/R_{\mathrm{max},0}$ begins to increase. This is because the core expands significantly due to tidal heating which results in larger $R_{\mathrm{max}}$.

We next increase $\beta$ from $3$ to $4$, which produces a more rapid decrease of density at large radii---results are shown in Fig.~\ref{fig:tidal_track_GADGET2}. Such profiles have been widely used in modeling the density profile of spherical galaxies, see e.g.~\cite{Hernquist:1990be,1993MNRAS.265..250D,1994AJ....107..634T,1993MNRAS.262.1062S}. Compared to the cases with $\beta=3$ and the same $\gamma$, the cases with $\beta=4$ are less influenced by tidal effects. For example, if we compare the blue circles in Fig.~\ref{fig:tidal_track_GADGET2} with the blue circles in Fig.~\ref{fig:tidal_track_GADGET}, both having the same $R_\mathrm{p}/R_\mathrm{a}$, at the end of the simulation $V_{\mathrm{max}}/V_{\mathrm{max},0}$ and $R_{\mathrm{max}}/R_{\mathrm{max},0}$ change less relative to their initial values than in the case with $\beta=4$. However, the slope of the tidal track in the case with $\beta=4$ is steeper.

Finally, we also run two cases with a different value of $\alpha=2$ as shown in Fig.~\ref{fig:tidal_track_GADGET3}. The parameter $\alpha$ controls how smoothly the density profile slope transits from inner value to the value at larger radii. For this value of $\alpha$, we limit our simulations to cored profiles with $\gamma=0$ and $\beta=3, 4$. Such density profiles have been found to well describe the observed density profile of Galactic globular clusters that are unaffected by tidal effects~\cite{Carballo-Bello:2011dgb}.~\footnote{Our parameter $\beta$ corresponds to the parameter $\gamma$ in Ref.~\cite{Carballo-Bello:2011dgb}.}

We fit the tidal tracks obtained from our simulations using the same fitting function introduced in Ref.~\cite{Penarrubia:2010jk}:
\begin{equation}
g(x)=\frac{2^{\mu} x^{\eta}}{(1+x)^{\mu}},
\label{eq:fitting}
\end{equation}
where $x=M_{\rm bound}(t)/M_{\rm bound}(0)$ is the bound mass fraction, and $g$ is $V_{\rm max}/V_{\rm max,0}$ or $R_{\rm max}/R_{\rm max,0}$. To mitigate the effect of numerical artifacts, we limit our fitting to the data with bound mass fraction $M_{\rm bound}/M_{\rm bound,0} > 10^{-3}$, so that the subhalo is resolved by more than $10^4$ particles. The best-fit parameters are listed in Table \ref{tab:fitting}. The fitting function with the best-fit parameters is shown in Figs.~\ref{fig:tidal_track_GADGET}, \ref{fig:tidal_track_GADGET2} and \ref{fig:tidal_track_GADGET3} by the solid lines.

\begin{table}
\caption{Best-fit parameters for Eq.~(\ref{eq:fitting}) for tidal tracks with different initial density profiles.}
\label{tab:fitting}
\begin{tabular}{c|c|c|r@{.}l|r@{.}l|r@{.}l|r@{.}l}
\hline
\boldmath{$\alpha$}      & \boldmath{$\beta$} & \boldmath{$\gamma$} & \multicolumn{2}{c}{\boldmath{$\mu_V$}}  & \multicolumn{2}{c}{\boldmath{$\eta_V$}}  & \multicolumn{2}{c}{\boldmath{$\mu_R$}}  & \multicolumn{2}{c}{\boldmath{$\eta_R$}}     \\
\hline
1 & 3 & 0~~~ & $0$&$8317$ & $0$&$4218$ & $-0$&$3737$   & $0$&$1976$    \\
1 & 3 & 0.5  & $0$&$7152$ & $0$&$3600$ & $-0$&$08328$  & $0$&$2819$    \\
1 & 3 & 1~~~ & $0$&$6175$ & $0$&$2895$ & $0$&$5529$    & $0$&$4675$    \\
1 & 3 & 1.5  & $0$&$3358$ & $0$&$1692$ & $1$&$207$     & $0$&$6845$    \\
\hline
1 & 4 & 0~~~ & $0$&$9149$ & $0$&$4982$ & $-0$&$1739$   & $0$&$1543$    \\
1 & 4 & 0.5  & $0$&$6286$ & $0$&$3968$ & $-0$&$003293$ & $0$&$2660$    \\
1 & 4 & 1~~~ & $0$&$4830$ & $0$&$3055$ & $0$&$5597$    & $0$&$4802$    \\
1 & 4 & 1.5  & $0$&$3469$ & $0$&$2018$ & $1$&$254$     & $0$&$7847$    \\
\hline
2 & 3 & 0~~~ & $1$&$055$  & $0$&$5114$ & $-0$&$06394$  & $0$&$1182$    \\
\hline
2 & 4 & 0~~~ & $1$&$412$  & $0$&$8288$ & $-0$&$1514$   & $-0$&$002544$ \\
\hline
\end{tabular}
\end{table}

For the case $\alpha=1$, $\beta=3$, and $\gamma=1$, Ref.~\cite{Errani:2020wgn} proposed a slightly different fitting formula:
\begin{equation}
V_{\rm max}/V_{\rm max,0}=\frac{2^{\alpha}\left(R_{\rm max}/R_{\rm max,0}\right)^{\beta}}{\left[1+\left(R_{\rm max}/R_{\rm max,0}\right)^2\right]^{-\alpha}}
\label{eq:fitting_Errani}
\end{equation}
with $\alpha=0.4$ and $\beta=0.65$. The best-fit function for this case found in this work is very close to the results of Ref.~\cite{Errani:2020wgn}.

For the case with $(\alpha,\beta,\gamma)=(1,3,0)$, the fitting function Eq.~\eqref{eq:fitting} cannot capture the turnaround of the tidal track (see the top left panel of Fig.~\ref{fig:tidal_track_GADGET}), so we exclude the last data point in the lower right corner from our fitting process.

\subsubsection{Convergence tests for extremely cuspy subhalos}\label{subsec:convergence}

\begin{figure*}
\includegraphics[width=0.95\columnwidth]{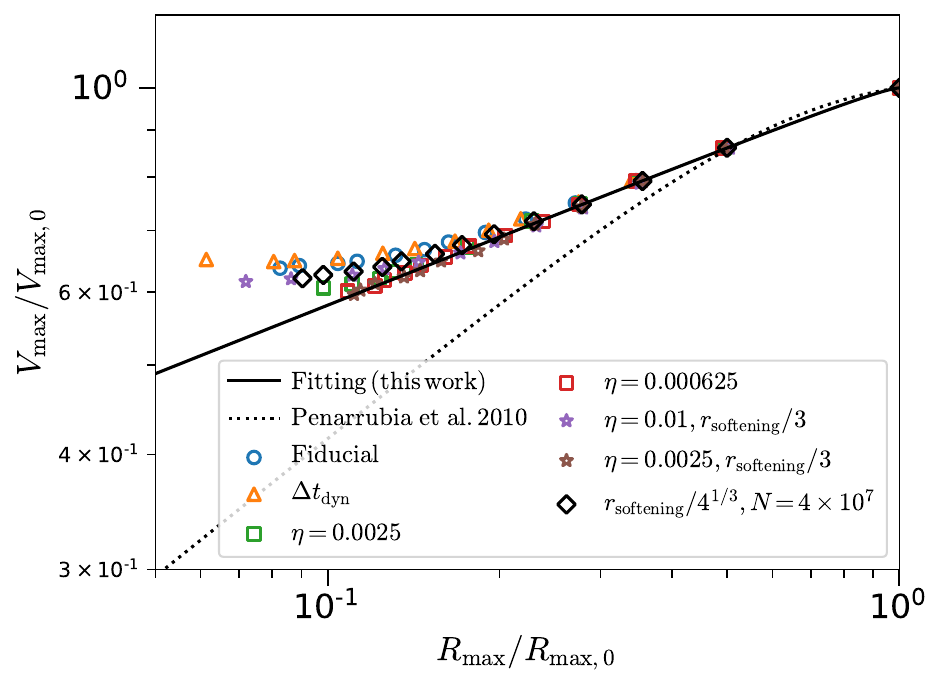}
\includegraphics[width=0.95\columnwidth]{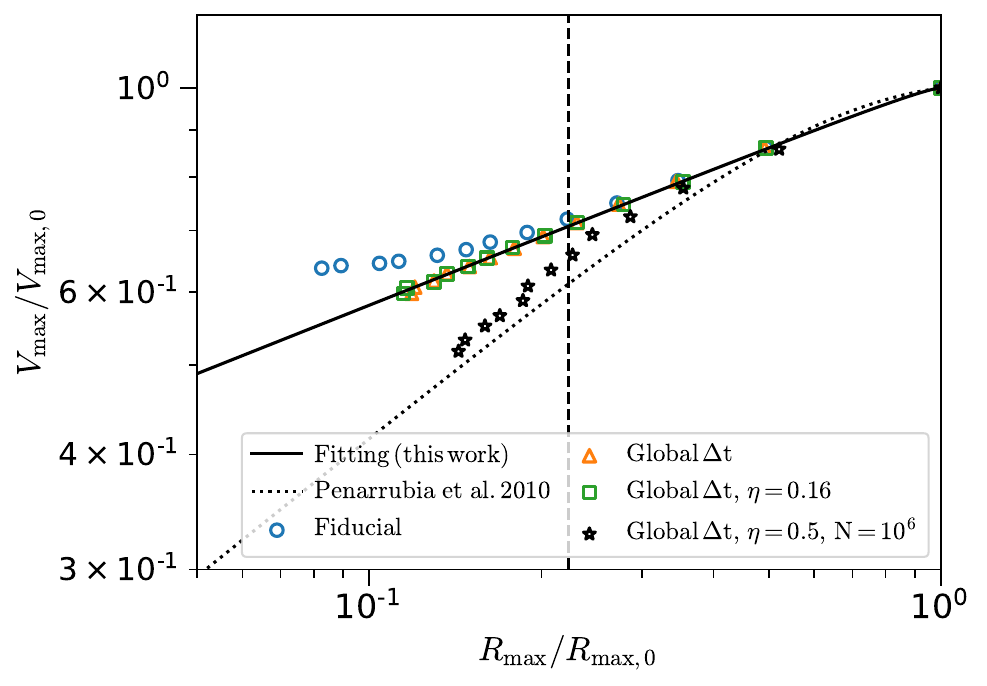}
\caption{Convergence test for $(\alpha,\beta,\gamma)=(1,3,1.5)$, $R_\mathrm{p}/R_\mathrm{a}=1/200$. Left: tidal tracks from $N$-body simulations with different time step size parameter $\eta$ (circle and squares) and softening length $r_{\rm softening}$ (stars). Triangles show the simulation results with an additional time step size criterion based on the local dynamical time scale, Eq.~\eqref{eq:dt_dyn}. Increasing the number of particles by a factor of $4$ and decreasing $r_{\rm softening}$ accordingly by a factor of $4^{1/3}$ (diamonds) do not grantee converged results. Right: tidal tracks from $N$-body simulations using a global time step size and different $\eta$. With a global time step size, converged results are obtained at larger $\eta$ than the cases using the fiducial adaptive time step size. For very large time steps, e.g. $\eta=0.5$ (black stars) results are not converged.}
\label{fig:tidal_track_convergence}
\end{figure*}

As our result for $\gamma=1.5$ differs significantly from that shown found by Ref.~\cite{Penarrubia:2010jk}, we conducted additional convergence tests for this case. There are three parameters that control the time and spatial resolution of the simulation: (1) time step size criterion; (2) softening length, $r_{\rm softening}$; (3) particle number, $N$.

In {\sc Gadget-4}, the time step size is chosen based on the softening length and the particle acceleration $a$, as defined in Eq.~\eqref{eq:time_step}.
Another time step size criterion that has been used in literature is based on the local dynamical time:
\begin{equation}
\Delta t = \frac{\eta_{\rm dyn}}{\sqrt{G \rho}},
\label{eq:dt_dyn}
\end{equation}
where $\rho$ is the local density. In this work $\eta_{\rm dyn}$ is set to the same value as the $\eta$ parameter used in the acceleration-based time step criterion, Eq.~\eqref{eq:time_step}. To compute the local density, we use the same algorithm as the smoothed-particle hydrodynamics (SPH) method implemented in {\sc Gadget-4} for gas particles~\cite{Springel:2020plp}, i.e. a cubic spline kernel is used to compute the density at the particle position from the $64$ nearest neighbor particles. Adding the local dynamical time step size criterion does not change the results too much, see the left panel of Fig.~\ref{fig:tidal_track_convergence}.

As we decrease the parameter $\eta$, the results slowly converge,
but the differences from the results of Ref.~\cite{Penarrubia:2010jk} remain. Possible explanations for the differences include the following: (i) To run the simulations, \cite{Penarrubia:2010jk} uses {\sc SUPERBOX} \cite{2000NewA....5..305F} which computes the gravitational interactions using multiple layers of grids while we use {\sc Gadget-4}, which computes the gravitational interactions using FFM; (ii) the details of our self-binding analysis differ from the approach in Ref.~\cite{Penarrubia:2010jk}.~\footnote{Even for the cases where the tidal tracks we find agree well with the results of \cite{Penarrubia:2010jk}, the dependence of $V_{\rm max}/V_{\rm max,0}$ and $R_{\rm max}/R_{\rm max,0}$ on $M_{\rm bound}/M_{\rm bound,0}$ differ from those found by \cite{Penarrubia:2010jk}, indicating some difference in which particles are considered to be bound to the subhalo. For example, for the case $(\alpha,\beta,\gamma)=(1,3,1)$, we find that we must multiply the bound mass we obtain by a factor of $2$, to obtain an approximate match with $V_{\mathrm{max}}$ as a function of $M_{\mathrm{bound}}/M_{\mathrm{bound},0}$ as found by \cite{Penarrubia:2010jk}, see Appendix~\ref{appendix:track_more}} 

Interestingly, we find that using a global time step size, i.e.\ all particles are evolved using the same time step, significantly improves convergence allowing the use of a larger time step, see the right panel of Fig.~\ref{fig:tidal_track_convergence}. The computing time required to achieve converged results is then much shorter than that using individual time step sizes for each particle. This suggests that to simulate extremely cuspy subhalos, the time step size criterion should be chosen carefully and using a global time step size may be helpful (see also Refs.~\cite{Fischer:2024eaz,Mace:2024}). We also find that, when we use a very large time step size, e.g. $\eta=0.5$, the results are unconverged, but are in fact closer to the fitting function obtained by \cite{Penarrubia:2010jk}. For this case, the time step size $\Delta t = 1.2\ {\mathrm{Myr}}$, which is still smaller than the one used by \cite{Penarrubia:2010jk}, i.e. $\Delta t = 4.6\ {\mathrm{Myr}}$. Therefore, it is possible that the results for the case with $\gamma=1.5$ shown in \cite{Penarrubia:2010jk} are not fully converged. A direct comparison between {\sc Gadget-4} and {\sc SUPERBOX} using the same initial conditions and post-analysis would be required to further confirm this hypothesis.

\subsection{Evolution of density profiles}\label{subsec:rho_evo}

In previous studies, Hayashi {\it et al.}~\cite{Hayashi:2002qv} and Pe\~narrubia {\it et al.}~\cite{Penarrubia:2007zx,Penarrubia:2010jk} have shown that, beyond just $V_{\mathrm{max}}$ and $R_{\mathrm{max}}$, the evolution of subhalo density profiles also follows a universal behavior and depends only on the remaining fraction of bound mass. A transfer function that connects the current density profile $\rho(r,t)$ to the initial one, $\rho(r,0)$, can be defined as
\begin{equation}
H(r, x) = \frac{\rho(r, t)}{\rho(r,0)},
\end{equation}
where $x=M_{\mathrm{bound}}/M_{\mathrm{bound}}(0)$ is the bound mass fraction.

In Ref.~\cite{Hayashi:2002qv}, a fitting function is proposed as
\begin{equation}
H(r, x) = \frac{f_\mathrm{t}}{1+(r/r_{\mathrm{te}})^{\delta}},
\label{eq:transfer}
\end{equation}
with $r_{\mathrm{te}}$ the effective tidal radius, $f_\mathrm{t}$ a normalization factor that quantifies the density drop in the center, and $\delta=3$. By calibrating to $N$-body simulations simulations, Ref.~\cite{Hayashi:2002qv} gives the following fitting formulae for $r_{\mathrm{te}}$ and $f_\mathrm{t}$:
\begin{eqnarray}
\log_{10} \frac{r_{\mathrm{te}}}{r_s} &=& 1.02 + 1.38 \log_{10} x + 0.37 (\log_{10} x)^2,
\label{eq:rte}\\
\log_{10} f_\mathrm{t} &=&0.007+ 0.35 \log_{10} x + 0.39 (\log_{10} x)^2 \nonumber \\
&&+ 0.23 (\log_{10} x)^3.
\label{eq:rft}
\end{eqnarray}

In a more recent study, Green {\it et al.}~\cite{Green:2019zkz} have examined the density evolution of subhalos utilizing the DASH library~\cite{Ogiya:2019del}, a large set of high-resolution, idealized simulations. They find that the density transfer function has an additional dependence on the initial concentration of subhalos and proposed a more general formula:
\begin{equation}
H(r, x, c) = \frac{f_\mathrm{t}}{1+\left[\frac{r}{r_{\mathrm{te}}}\left(\frac{r_{\mathrm{vir,sub}}-r_{\mathrm{te}}}{r_{\mathrm{vir,sub}} }\right)\right]^{\delta}},
\label{eq:transfer_green}
\end{equation}
where the concentration parameter $c$ enters in the virial radius, and $r_{\mathrm{te}}$, $f_\mathrm{t}$ and $\delta$ are also concentration dependent:
\begin{eqnarray}
r_{\mathrm{te}}&=&r_{\mathrm{vir,sub}} x^{b_1 (c/10)^{b_2}} c^{b_3(1-x)^{b_4}} \nonumber \\
&&\exp\left[b_5\left(\frac{c}{10}\right)^{b_6}(1-x)\right],
\label{eq:rte_green}\\
f_\mathrm{t}&=&x^{a_1 (c/10)^{a_2}} c^{a_3(1-x)^{a_4}},\\
\delta &=& c_0\,x^{c_1 (c/10)^{c_2}}c^{c_3(1-x)^{c_4}}.
\label{eq:delta_green}
\end{eqnarray}
The values of the parameters in the above equations are given in Table 1 of Ref.~\cite{Green:2019zkz}.

Since, in the current work, we fix the subhalo concentration at $20.5$ [see Eqs.~\eqref{eq:rs} and \eqref{eq:rvir}], we choose to fit the density profile measured from our simulations using the simpler formula, Eq.~\eqref{eq:transfer}. Reference~\cite{Green:2019zkz} found that their best fit value for $\delta$ is 2--3. Therefore, we consider two cases: $\delta=3$ and $\delta=2$. We emphasize that sometimes it can be useful to have a density profile that leads to analytic formula for certain physical quantities, such as the enclosed mass within a given radius. The case with $\delta=2$ has been widely used in modeling subhalo density profiles in the analysis of strong gravitational lenses (e.g.~\cite{Gilman:2019nap}). Applying the transfer function Eq.~\eqref{eq:transfer} with $\delta=2$ to the NFW profile, the lensing convergence can be computed analytically~\cite{Baltz:2007vq}.

\begin{figure*}
\includegraphics[width=\columnwidth]{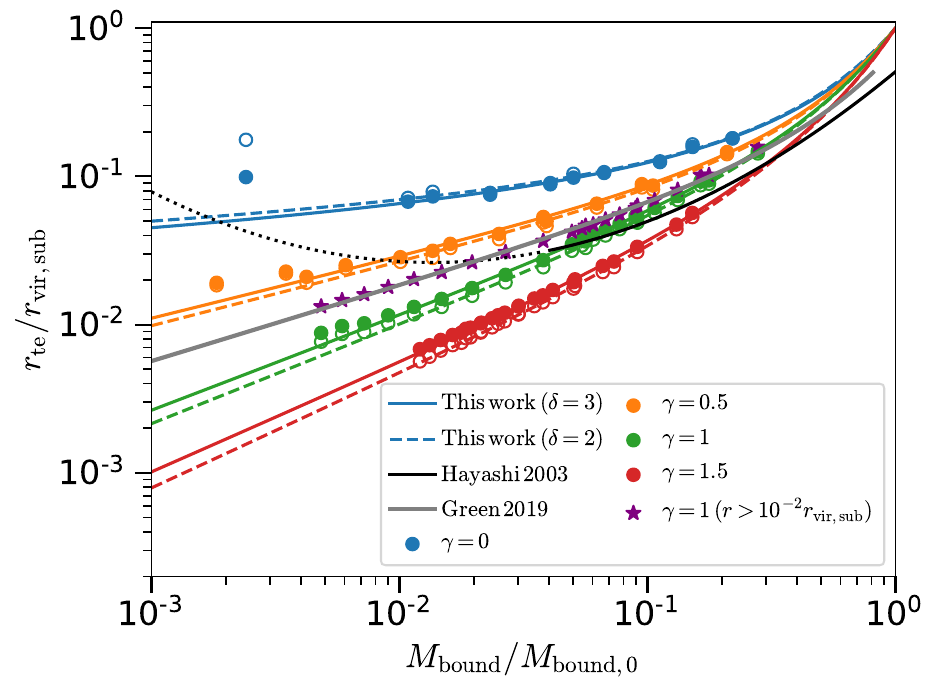}
\includegraphics[width=\columnwidth]{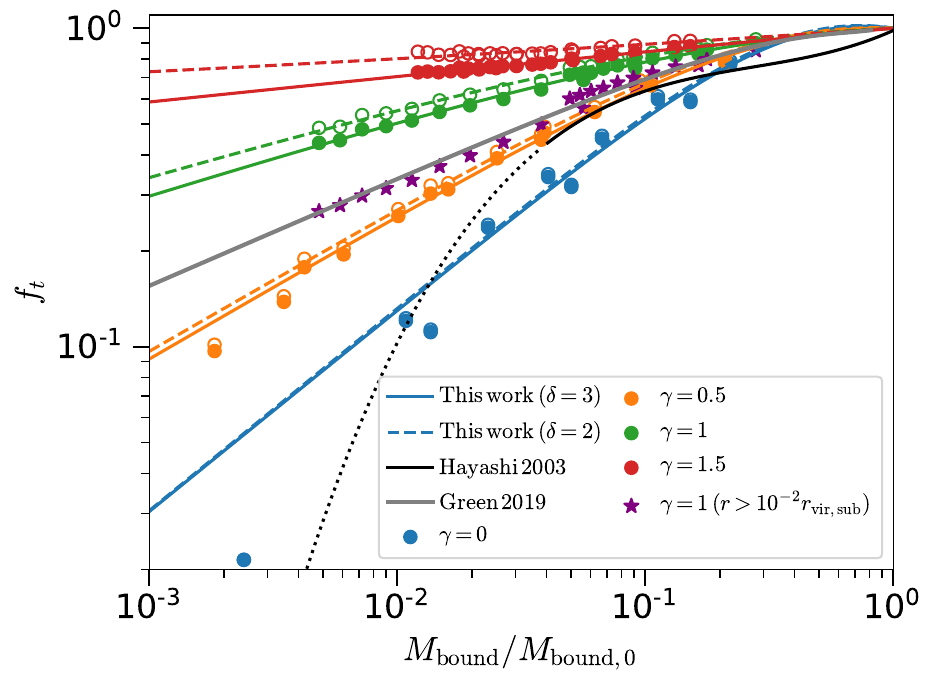}
\caption{Best-fit parameters for the density profile as a function of bound mass fraction for $(\alpha,\beta)=(1,3)$ and different $\gamma$. Left: the effective tidal radius in Eq.~\eqref{eq:transfer}. The filled (open) circles are obtained by fitting Eq.~\eqref{eq:transfer} to the density profiles measured from simulation assuming $\delta=3$ ($\delta=2$). For each $\gamma$, data from simulations with different $R_\mathrm{p}/R_\mathrm{a}$ are combined. The colored lines show the fitting formula Eq.~\eqref{eq:rte_new} with our best-fit parameter values. The fitting functions from Hayashi {\it et al.} 2003~\cite{Hayashi:2002qv} (black) and Green {\it et al.} 2019~\cite{Green:2019zkz} (gray) are also shown. Right: the normalization parameter $f_\mathrm{t}$ in Eq~\eqref{eq:transfer}. The purple stars show the best-fit values for $r_{\mathrm{te}}$ and $f_\mathrm{t}$ when excluding measurements of the density profiles at $r<10^{-2}$. The Hayashi {\it et al.} 2003 fitting function is calibrated only for $M_{\mathrm{bound}}/M_{\mathrm{bound,0}}>4\times10^{-2}$, thus at small mass ratios an extrapolation has be used (black dotted line).}
\label{fig:den_fit}
\end{figure*}

We first measure the density transfer function $H$ of subhalos at the apocenters of their orbits from our simulations with different initial subhalo profiles. Then we fit Eq.~\eqref{eq:transfer} to these measured transfer functions to find the best fit $r_{\mathrm{te}}$ and $r_t$. $r_{\mathrm{te}}$ and $r_t$ as a function of the bound mass fraction for different $\gamma$. Results are shown in Fig.~\ref{fig:den_fit} (colored markers). For each value of $\gamma$, we have combined the simulation data for different subhalo orbits, i.e. different value of $R_\mathrm{p}/R_\mathrm{a}$. Inspired by the work of Refs.~\cite{Hayashi:2002qv,Green:2019zkz}, we fit the mass dependence in $r_{\mathrm{te}}$ and $r_t$ assuming
\begin{eqnarray}
\frac{r_{\mathrm{te}}}{r_{\mathrm{vir,sub}}}&=&\frac{(1+A)\,x^B}{1+A\,x^{2 B}}\exp\left[-C (1-x)\right],
\label{eq:rte_new}\\
f_{t} &=& \frac{(1+D)\,x^E}{1+D\,x^{2 E}}.
\label{eq:ft_new}
\end{eqnarray}
These functions are chosen to ensure that at the beginning of the simulation ($x=1$), $r_{\mathrm{te}}=r_{\mathrm{vir,sub}}$ and $f_\mathrm{t}=1$.

The best fit formula are shown in Fig.~\ref{fig:den_fit} (colored curves) for $(\alpha,\beta)=(1,3)$ and different $\gamma$. We also compare our fitting functions to those obtained by Hayashi {\it et al.}~\cite{Hayashi:2002qv} (black curve) and Green {\it et al.}~\cite{Green:2019zkz} (gray curve) for initial NFW profiles, i.e. $\gamma=1$. We find that the effective tidal radius $r_{\mathrm{te}}$ we obtain is in broad agreement with that from Hayashi {\it et al.} at $x > 4\times 10^{-2}$. At $x < 4\times 10^{-2}$, the fitting formula from Hayashi {\it et al.} shows an upturn, which does not appear in our results. However, we note that the Hayashi {\it et al.} fitting function was calibrated only for $x > 4\times 10^{-2}$. Our result for $r_{\mathrm{te}}$ is lower than that found by Green {\it et al.} Here we have taken into account the difference in definition of halo concentration in our work from Green {\it et al.}, but this has only a small effect. Since the number of particles in our simulations is $10$ times higher than that used by Green {\it et al.}, we can better resolve the central region of the subhalo. We have tried excluding radii bins with $r < 10^{-2} r_{\mathrm{vir,sub}}$ from our fitting. Our results are then in good agreement with those from Green {\it et al.}, as shown by the stars in Fig.~\ref{fig:den_fit}. This suggests that the difference we see is due to numerical resolution and the range of radii bins used in the fitting process. For $f_\mathrm{t}$, a similar difference between our results and those obtained by Green {\it et al.} is also seen, as shown in the right panel of Fig.~\ref{fig:den_fit}. The fitting function from Green {\it et al.} overestimates the density decrease in the central region of subhalos due to tidal effects. Similar findings has been reported in Ref.~\cite{Errani:2020wgn}. Again, we show that we can recover the Green {\it et al.} results by excluding the data points at small radii from the our fitting procedure.

\begin{figure*}
\includegraphics[width=\columnwidth]{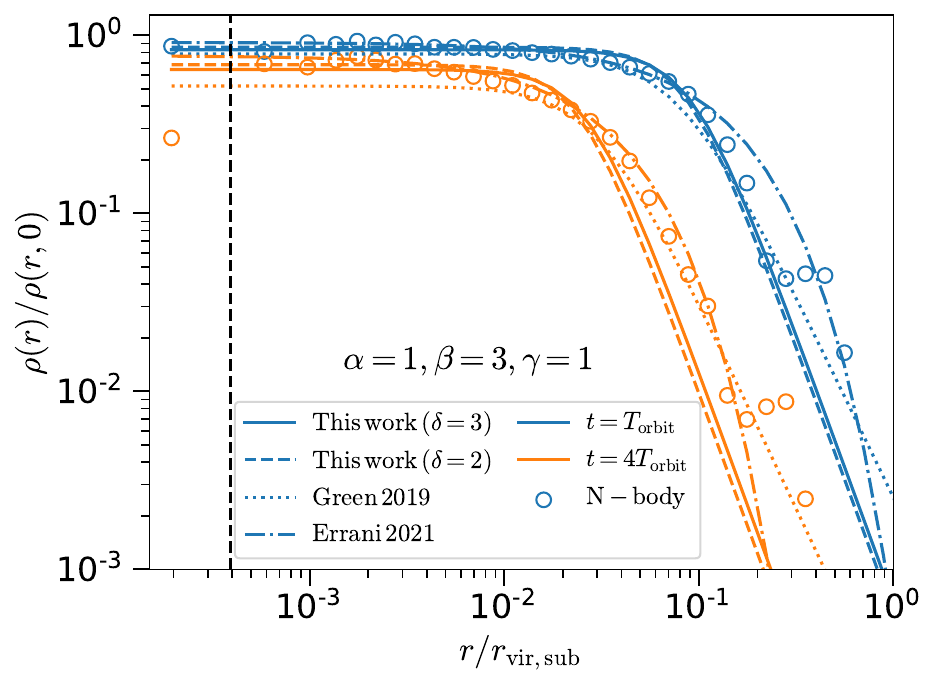}
\includegraphics[width=\columnwidth]{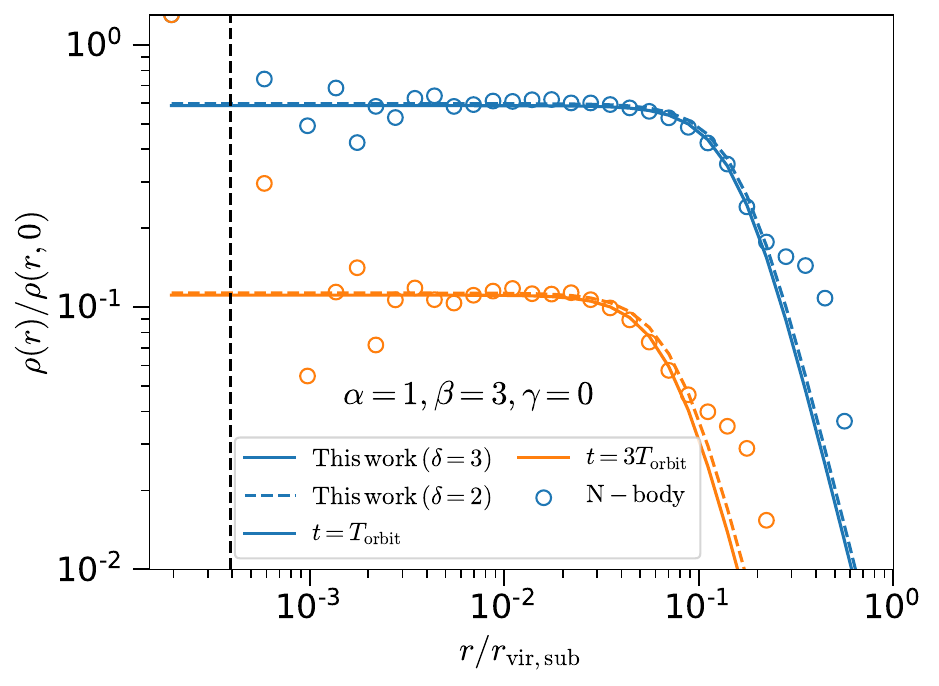}
\caption{Density profiles of subhalos from the fitting functions (lines; with line type indicating the specific fitting function used as indicated in the panel) compared with the simulation data (colored circles). Left: the subhalo initially has a cuspy core with $(\alpha,\beta,\gamma)=(1,3,1)$. The blue (orange) circles shows the density profiles after the subhalo completes 1 (4) orbit(s) as indicated in the panel. Right: the subhalo initially has a flat core with $(\alpha,\beta,\gamma)=(1,3,0)$. The vertical dashed lines indicate $2.8 \,r_{\mathrm{softening}}$.}
\label{fig:den_compare}
\end{figure*}

In Fig.~\ref{fig:den_compare}, we show the density transfer function of subhalos at different times for cuspy (left panel) and cored (right panel) initial profiles. For comparison, we also show our best fit formula and the models from Green {\it et al.}~\cite{Green:2019zkz} and Errani {\it et al.}~\cite{Errani:2020wgn}.

For the cuspy case, after one orbit we see that our fitting results (solid and dashed curves) better capture the density suppression at large radii than does the model by Errani {\it et al.} After $4$ orbits, when the subhalo is heavily stripped, our fitting results perform slightly worse that those from Errani {\it et al.}, but are still in good agreement with the simulation data (open circles) below $0.2 r_{\mathrm{vir,sub}}$. On the other hand, the model from Green {\it et al.}\ underestimates the central density of subhalos when subhalos are heavily stripped.

For the cored case, ours fitting results also work reasonably well, even while cored subhalos are more strongly influenced by tidal effects. After $3$ orbits, the central density of the subhalo decreases to $10\%$ of its initial value.

\begin{table}
\caption{Best-fit parameters in Eqs.~\eqref{eq:rte_new} and \eqref{eq:ft_new} for density transfer function Eq.~\eqref{eq:transfer} with different initial profiles. $\delta=3$ has been assumed.}
\label{tab:fitting_density_1}
\begin{tabular}{c|c|c|r@{.}l|r@{.}l|r@{.}l|r@{.}l|r@{.}l}
\hline
\boldmath{$\alpha$}  & \boldmath{$\beta$}   & \boldmath{$\gamma$}      & \multicolumn{2}{c}{\boldmath{$A$}}  & \multicolumn{2}{c}{\boldmath{$B$}} & \multicolumn{2}{c}{\boldmath{$C$}}     & \multicolumn{2}{c}{\boldmath{$D$}} & \multicolumn{2}{c}{\boldmath{$E$}}     \\
\hline
1 & 3 & 0~~~ & $-0$&$9309$  & $0$&$04703$ & $0$&$7684$  & $1$&$402$   & $0$&$6325$   \\
1 & 3 & 0.5  & $-0$&$2291$  & $0$&$4123$  & $1$&$399$   & $1$&$087$   & $0$&$4523$   \\
1 & 3 & 1~~~ & $0$&$9093$   & $0$&$6368$  & $2$&$185$   & $1$&$436$   & $-0$&$2491$  \\
1 & 3 & 1.5  & $0$&$8353$   & $0$&$7340$  & $2$&$432$   & $0$&$08093$ & $0$&$08491$  \\
\hline
1 & 4 & 0~~~ & $41$&$33$    & $0$&$6082$  & $4$&$070$   & $1$&$088$   & $0$&$7280$   \\
1 & 4 & 0.5  & $0$&$4212$   & $-0$&$3816$ & $3$&$629$   & $0$&$6879$  & $0$&$4862$   \\
1 & 4 & 1~~~ & $15$&$92$    & $0$&$7194$  & $4$&$982$   & $0$&$3359$  & $0$&$2508$   \\
1 & 4 & 1.5  & $104$&$1$    & $1$&$247$   & $6$&$630$   & $0$&$1514$  & $0$&$1284$   \\
\hline
2 & 3 & 0~~~ & $-0$&$9646$  & $-0$&$002371$ & $2$&$511$  & $1$&$314$  & $0$&$8240$   \\
\hline
2 & 4 & 0~~~ & $5$&$793$E-7 & $-1$&$623$    & $5$&$471$  & $0$&$7780$ & $1$&$213$    \\
\hline
\end{tabular}
\end{table}

\begin{table}
\caption{Best-fit parameters in Eqs.~\eqref{eq:rte_new} and \eqref{eq:ft_new} for density transfer function Eq.~\eqref{eq:transfer} with different initial profiles. $\delta=2$ has been assumed.}
\label{tab:fitting_density_2}
\begin{tabular}{c|c|c|r@{.}l|r@{.}l|r@{.}l|r@{.}l|r@{.}l}
\hline
\boldmath{$\alpha$}   & \boldmath{$\beta$}   & \boldmath{$\gamma$}      & \multicolumn{2}{c}{\boldmath{$A$}}  & \multicolumn{2}{c}{\boldmath{$B$}} & \multicolumn{2}{c}{\boldmath{$C$}}     & \multicolumn{2}{c}{\boldmath{$D$}} & \multicolumn{2}{c}{\boldmath{$E$}}     \\
\hline
1 & 3 & 0~~~ & $-0$&$9984$  & $5$&$503$E-4 & $1$&$235$   & $1$&$482$ & $0$&$6364$     \\
1 & 3 & 0.5  & $-0$&$1921$  & $0$&$4268$   & $1$&$463$   & $1$&$171$ & $0$&$4500$     \\
1 & 3 & 1~~~ & $0$&$6849$   & $0$&$6644$   & $2$&$078$   & $0$&$7583$ & $0$&$2338$    \\
1 & 3 & 1.5  & $0$&$9839$   & $0$&$7688$   & $2$&$518$   & $-0$&$9951$ & $1$&$315$E-4 \\
\hline
1 & 4 & 0~~~ & $39$&$42$    & $0$&$5883$   & $4$&$054$   & $1$&$185$ & $0$&$7373$    \\
1 & 4 & 0.5  & $2$&$066$    & $0$&$3871$   & $3$&$636$   & $0$&$7524$ & $0$&$4777$   \\
1 & 4 & 1~~~ & $13$&$99$    & $0$&$7222$   & $5$&$009$   & $0$&$3684$ & $0$&$2231$   \\
1 & 4 & 1.5  & $83$&$40$    & $1$&$216$    & $6$&$620$   & $0$&$2850$ & $0$&$1053$   \\
\hline
2 & 3 & 0~~~ & $-1$&$024$   & $0$&$001643$ & $2$&$413$   & $1$&$371$  & $0$&$8315$   \\
\hline
2 & 4 & 0~~~ & $3$&$273$E-6 & $-1$&$946$   & $5$&$847$   & $0$&$8384$ & $1$&$231$    \\
\hline
\end{tabular}
\end{table}

A full list of the best fit parameters for subhalos with different combinations of $(\alpha,\beta,\gamma)$ is shown in Tables~\ref{tab:fitting_density_1} and \ref{tab:fitting_density_2}. We note that the fitting functions Eqs.~\eqref{eq:rte_new} and \eqref{eq:ft_new} we choose result in tidal radii and central densities that decrease as a power-law in the bound mass fraction at very small bound mass fractions. This is a good assumption in most cases, but, for the case with a flat core, i.e. $\gamma=0$, tidal heating can lead to significant core expansion (see Sec.~\ref{sec:sams}), which makes $r_{\mathrm{te}}$ eventually begin to increase with decreasing bound mass faction (see the left panels of Figs.~\ref{fig:den_fit} and \ref{fig:den_fit_more_2}). This feature can be partially captured by our fitting functions, but the fitting function will drop again at sufficiently low $x$, see Appendix~\ref{appendix:den_fit_more}. For this reason, the best fit formula for the cases with $\gamma=0$ should be used with caution at a bound mass fraction much smaller than the value where we have available measurements.

\section{Semi-analytic models}\label{sec:sams}

In the previous section, we showed tidal tracks and fitting formula results for DM subhalos with different initial density profiles. These empirical fitting formulas can be implemented in semi-analytic models such as {\sc Galacticus} \cite{2012NewA...17..175B} to study the statistical properties of subhalos. On the other hand, a physical model for tidal evolution that can reproduce these tidal tracks would be extremely useful, especially when extrapolating tidal tracks into regimes where artificial disruption~\cite{2018MNRAS.474.3043V} can occur. In a previous study~\cite{2022MNRAS.517.1398B}, we showed that by using an improved model for tidal heating that accounts for the second-order heating terms, we could accurately reproduce the tidal track for NFW profiles. In this work, we will extend the model of Ref.~\cite{2022MNRAS.517.1398B} to the more general density profile in Eq.~\eqref{eq:prof_abg}. Furthermore, in many applications, in addition to the $V_{\mathrm{max}}$--$R_{\mathrm{max}}$ track, the $V_{\mathrm{max}}$--$M_{\rm bound}$ relation, and the time evolution of $M_{\rm bound}$ are required in order to build a complete model for the evolution of subhalos. In Ref.~\cite{2020MNRAS.498.3902Y}, the tidal stripping model in {\sc Galacticus} was calibrated to cosmological cold dark matter $N$-body simulations, ELVIS \cite{Garrison-Kimmel:2013eoa} and Caterpillar \cite{Griffen:2015rqk}. In this work, we extend this tidal stripping model and recalibrate it to our simulations of subhalos with different density profiles.

In the remainder of this section we detail the orbital and tidal physics included in our model, and present results for the calibration of the parameters of this model. All the models described below are implemented in the public semi-analytic model for galaxy formation, Galacticus \cite{2012NewA...17..175B}.~\footnote{\url{https://github.com/galacticusorg/galacticus}}

\subsection{Orbital evolution}\label{subsec:orbit}
When a subhalo evolves in a host potential, its acceleration can be written as
\begin{equation}
    \mathbf{a}=\mathbf{a}_\mathrm{g} + \mathbf{a}_\mathrm{df},
\end{equation}
where $\mathbf{a}_\mathrm{g}$ is the gravitational acceleration from the host, and $\mathbf{a}_\mathrm{df}$ is the dynamical friction caused by the overdense wake of host particles that generated behind the subhalo when it orbits within the host. Using the Chandrasekhar formula, $\mathbf{a}_\mathrm{df}$ can be computed as~\cite{1943ApJ....97..255C}
\begin{eqnarray}
\mathbf{a}_{\rm{df}}=-4\pi \mathrm{G}^2 \ln\Lambda M_{\rm{sub}}\rho_{\rm{host}}(r_{\rm{sub}})\dfrac{\mathbf{V}_{\rm{sub}}}{V_{\rm{sub}}^3}\nonumber \\
\left[{\rm erf}(X_v)-\dfrac{2X_v}{\sqrt{\pi}}\exp\left(-X_v^2\right)\right],
\label{eq:dyn_friction}
\end{eqnarray}
where $\mathrm{G}$ is the gravitational constant, $M_{\rm sub}$ is the bound mass of the subhalo, $\rho_{\rm host}$ is the host density at the subhalo position, $r_{\rm sub}$ is the distance to the host center, $X_v=V_{\rm{sub}} / \sqrt{2}\sigma_v$ with $V_{\rm{sub}}$ the velocity of subhalo, $\sigma_v$ is the velocity dispersion of host particles, and $\ln\Lambda$ is the Coulomb logarithm.

Dynamical friction is only significant for subhalos with large mass ratios $M_{\rm sub}/M_{\rm host}$. In our simulations, $M_{\rm sub}/M_{\rm host}=1/1000$, and the dynamical friction effect is not relevant. Furthermore, we treat the host as a static potential, which will not generate dynamical friction since the host does not respond to the gravity of the subhalo. However, the subhalo's orbital radius still decays slowly with time due to the so-called ``self-friction'' effect \citep{2006PASJ...58..743F,2007MNRAS.375..604F,vandenBosch:2018tyt,Ogiya:2019del,2020MNRAS.495.4496M}, which arises from the interaction between the subhalo and particles stripped away from it through tidal forces. A detailed treatment of self-friction will be presented elsewhere~\cite{Du:2024}. In the current work, we mimic this effect approximately using \eqref{eq:dyn_friction} and adjust $\ln\Lambda$ to match the orbital evolution of subhalos measured from simulations. The details of this treatment do not significantly affect the calibration of our model.

\subsection{Tidal stripping}\label{subsec:stripping}

Subhalos are subject to the tidal force from the host. The tidal force pulls material in the subhalo away from its center. When the gravitational attraction from the subhalo is smaller than the tidal force, the subhalo particles will be able to become unbound, leading to mass loss from the subhalo. This happens outside the tidal radius which is defined as
\begin{equation}
    r_\mathrm{t}=\left(\dfrac{G M_{\mathrm{sub}}(<r_\mathrm{t})}{\gamma_\mathrm{c}\omega^2-\left.\frac{\mathrm{d}^2\Phi}{\mathrm{d}r^2}\right\vert_{r_{\mathrm{sub}}}}\right)^{1/3}.
\label{eq:r_t}
\end{equation}
Here $\omega=|\mathbf{r}_{\rm sub}\times\mathbf{V}_{\rm sub}|/\mathbf{V}_{\rm sub}^2$ is the angular frequency of the subhalo orbit, and $\Phi$ is the gravitational potential of the host at the subhalo position. The term $\gamma_\mathrm{c} \omega^2$ accounts for the centrifugal force in the coordinate system rotating with the subhalo. Different definitions of tidal radius have been used in previous studies, for example $\gamma_\mathrm{c}=0$ are used by Refs.~\cite{Tormen:1997ik,Klypin:1997fb,Hayashi:2002qv,Taffoni:2003sr}, and $\gamma_\mathrm{c}=1$ are used in Refs~\cite{1962AJ.....67..471K,2017MNRAS.471.4170T}. In Ref.~\cite{2018MNRAS.474.3043V}, both definitions have been tested against idealized simulations, and the authors found that neither case can perfectly reproduce the tidal mass loss measured in simulations. However, in their calculations, they did not take into account the change of subhalo density profile when it loses mass and is heated by tidal shocks. In the current work, we find that after accounting for the evolution of the density profile, as described in the next subsection, fixing $\gamma_\mathrm{c}=0$ gives a better match to simulation results.

Given the tidal radius, the subhalo mass outside of this is assumed to be lost on a timescale $T_{\rm loss}$:
\begin{equation}
    \frac{dM_{\rm sub}}{dt}=-\alpha_\mathrm{s} \frac{M_{\rm sub}-M_{\rm{sub}}(<r_\mathrm{t})}{T_{\rm loss}},
\end{equation}
where $\alpha_\mathrm{s}$ is a free parameter that controls the efficiency of tidal stripping.
There exist several relevant physical time scales that could be chosen for $T_\mathrm{loss}$: (i) orbital time scale: $T_{\rm orbit}=2\pi/\max\{\omega_\mathrm{t}, \omega_\mathrm{r}\}$ with $\omega_\mathrm{t}$ and $\omega_\mathrm{r}$ are the instantaneous frequencies of tangential and radial motion, respectively (this choice is adopted by \citealt{2020MNRAS.498.3902Y}); (ii) the dynamical time scale $T_{\rm dyn}=2\pi\sqrt{r_\mathrm{t}^3/16 \mathrm{G} M_{\rm sub}(<r_\mathrm{t})}$~\cite{2008gady.book.....B}. These two time scales are related but may differ significantly when the subhalo is close to the pericenter of its orbit. In this work, we find that the latter results in better fits to measurements from our simulations. Thus we will use it as our fiducial choice.

\subsection{Tidal heating}\label{subsec:heating}

As a subhalo orbits within its host and loses mass due to tidal stripping, its density profile also changes with time due to two effects: (i) after some mass is removed from the subhalo, it will revirialize and approach a new equilibrium with a different density profile; (ii) particles in the subhalo gain energy from tidal shocks, also known as the tidal heating effect, leading to the expansion of subhalo. When the subhalo passes through the pericenter of its orbit, both tidal stripping and tidal heating are strong. The subhalo loses a large fraction of the mass outside the tidal radius and at the same time is heated. As the subhalo approaches the apocenter of its orbit, it will begin to revirialize, resulting in a less concentrated density profile. As the subhalo becomes less concentrated, the tidal radius shrinks further and more mass will be stripped from the subhalo. The process of revirialization is complicated~\cite{Gnedin:1999rg}. In the current work, we focus on the density profile at successive apocenters for two reasons. First, the subhalo spends more time near apocenter during its orbit through its host. Second, at apocenter, the subhalos have had enough time since the strong mass loss and heating at pericenter to be revirialized, allowing us to apply the heating model proposed in~\cite{Taylor:2000zs,Pullen:2014gna}.

In these models of tidal heating, each spherical mass shell in the subhalo receives some heating energy resulting in a change in its specific energy changes of $\Delta \epsilon$. As a result, the mass shell expands from its initial radius $r_\mathrm{i}$ to a final radius $r_\mathrm{f}$ after revirialization. If no shell crossing happens, using the virial theorem and energy conservation, we have
\begin{equation}
\Delta \epsilon = \frac{\mathrm{G} M_\mathrm{i}}{2 r_\mathrm{i}}-\frac{\mathrm{G} M_\mathrm{i}}{2 r_\mathrm{f}},
\label{eq:r_i_r_f}
\end{equation}
where $M_\mathrm{i}$ is the enclosed mass within $r_\mathrm{i}$. Note that if no shell crossing happens, $M_\mathrm{i}$ is unchanged after expansion by definition. Given the initial density of the mass shell, $\rho_\mathrm{i}$, and the relation between $r_\mathrm{i}$ and $r_\mathrm{f}$ from solving Eq.~\eqref{eq:r_i_r_f}, the density of the mass shell after reaching new equilibrium can be written as
\begin{equation}
\rho_\mathrm{f} = \rho_\mathrm{i} \frac{r_\mathrm{i}^2}{r_\mathrm{f}^2} \frac{\mathrm{d}r_\mathrm{i}}{\mathrm{d}r_\mathrm{f}},
\label{eq:rho_f}
\end{equation}
where we have assumed mass conservation. Knowing the final  mass profile, we can then compute $V_{\mathrm{max}}$ and $R_{\mathrm{max}}$ and predict the tidal track for subhalos starting from a chosen initial profile.

To find $\Delta \epsilon$, we use the impulse approximation~\cite{Gnedin:1997vp,Gnedin:1999rg} and compute the heating rate per unit mass as~\cite{Taylor:2000zs}
\begin{equation}
\Delta\dot{\epsilon}(r) = \frac{\epsilon_\mathrm{h}}{3} \left[1+(\omega_\mathrm{p} T_{\rm shock})^2\right]^{-\gamma_\mathrm{h}} r^2 g_\mathrm{ab} G^\mathrm{ab},
\label{eq:heating_rate}
\end{equation}
where $\epsilon_\mathrm{h}$ is a coefficient needed to be calibrated to simulations, $\omega_\mathrm{p}$ is the angular frequency of particles at the half-mass radius of the subhalo, $T_{\rm shock}=r_{\rm sub}/V_{\rm sub}$ is the time scale of tidal shock, $\gamma_\mathrm{h}$ is the adiabatic index, $g_{ab}$ is the tidal tensor, and $G_{ab}$ is the time integral of $g_{ab}$ \cite{Pullen:2014gna,2020MNRAS.498.3902Y}:
\begin{equation}
G_{ab}(t)=\int_0^t dt' \left[ g_{ab}(t)-\beta_\mathrm{h}\frac{G_{ab}(t)}{T_{\rm orbit}}\right].
\label{eq:Gab}
\end{equation}
Here, for repeated indices, Einstein summation convention is adopted. As in \cite{2020MNRAS.498.3902Y}, a decaying term $-\frac{G_\mathrm{ab}}{T_{\rm orbit}}$ is added to the integral~\eqref{eq:Gab} to account for the fact that the impulse approximation is not valid on time scales larger than $T_{\rm orbit}$ when the movement of particles within the subhalo are non-negligible. In this work, we introduce a new coefficient $\beta_\mathrm{h}$ which controls the precise timescale for this decaying term.

The term in the square brackets in Eq.~\eqref{eq:heating_rate} accounts for the adiabatic correction, i.e. the heating energy gained by particles with orbital timescale much smaller than $T_{\rm shock}$ is suppressed due to ``adiabatic shielding"~\cite{SpitzerJr+1988,Weinberg:1994ux,Weinberg:1994uy}. Gnedin {\it et al.} ~\cite{Gnedin:1999rg} find $\gamma_\mathrm{h}=2.5$ for $\omega_\mathrm{p} T_{\rm shock}\lesssim 1$, while it is shown that $\gamma_\mathrm{h}=1.5$ in the regime $\omega_\mathrm{p} T_{\rm shock}\gtrsim 4$. On the other hand, \cite{2020MNRAS.498.3902Y} find that a value of $\gamma_\mathrm{h}=0$ predicts a $V_{\mathrm{max}}-M_{\mathrm{bound}}$ relation that is in better agreement with high-resolution cosmological $N$-body simulations Caterpillar \citep{Griffen:2015rqk} and ELVIS \citep{Garrison-Kimmel:2013eoa}. In this work, we have tried both $\gamma_\mathrm{h}=2.5$ and $\gamma_\mathrm{h}=0$, and reach a similar conclusion as \cite{2020MNRAS.498.3902Y}, i.e. $\gamma_\mathrm{h}=0$ results in $V_{\mathrm{max}}$ matching more closely with simulation results. This is partially due to the fact that the decaying term introduced in Eq.~\eqref{eq:Gab} also acts to suppress the heating energy.

As shown in Ref.~\cite{2022MNRAS.517.1398B}, using Eq.~\eqref{eq:heating_rate} to compute the heating energy results in a reasonable match to the tidal tracks of NFW subhalos. However, they find that to get a more accurate model, one needs to take into account the second-order energy perturbations $\langle E^2 \rangle^{1/2}$ explicitly. The second-order energy perturbation is usually of the same order as the first-order term given in Eq.~\eqref{eq:heating_rate}, but has a different radial dependence~\cite{Gnedin:1999rg,2022MNRAS.517.1398B}. Following \cite{2022MNRAS.517.1398B}, we write the total heating energy as
\begin{eqnarray}
\Delta \epsilon(r) &=& \Delta \epsilon_1 (r) + \Delta \epsilon_2 (r) \nonumber \\
&=& \Delta \epsilon_1 (r) + \sqrt{2} f_2 (1+\chi_v) \sqrt{\Delta \epsilon_1(r) \sigma^2_r(r)},
\label{eq:heating_improve}
\end{eqnarray}
where $\epsilon_1$ and $\epsilon_2$ are the contributions from the first-order and second-order energy perturbations, $f_2$ is a coefficient, $\chi_v$ is the position-velocity correlation (see Ref.~\cite{Gnedin:1999rg}), and $\sigma_r$ is the radial velocity dispersion of subhalo prior to any tidal heating. Reference~\cite{Gnedin:1999rg} shows that $\chi_v$ depends only weakly on the density profile, so we fix $\chi_v$ at a typical value of $-0.333$. Any uncertainty of $\chi_v$ is absorbed in the coefficient $f_2$.

In the heating model presented above, we assumed that there is no shell crossing. However, this may not be valid for all dark matter halos, especially those with cored density profiles. From Eqs.~\eqref{eq:heating_rate} and \eqref{eq:heating_improve}, we can see that the total heating energy $\Delta \epsilon(r)\propto r$ at small radii. For a density profile with an inner logarithmic slope of $-\gamma$, i.e. $\rho(r)\propto r^{-\gamma}$, the gravitational potential $\Phi(r)\sim \mathrm{G} M(r) / r \propto r^{2-\gamma}$. Thus if $\gamma < 1$ there always exist a radius $r_0$ below which $\Delta \epsilon(r) > \mathrm{G} M(r) / 2 r$. According to Eq.~\eqref{eq:r_i_r_f}, mass shells with $r_\mathrm{i} < r_0$ will then have a final radius of infinity, which means that the no shell crossing assumption is broken. More accurately, shell crossing happens when $\mathrm{d}r_\mathrm{f}/\mathrm{d}r_\mathrm{i} < 0$, in which case the enclosed mass with $r_\mathrm{i}$ is no longer constant and one needs to take into account this in Eq.~\eqref{eq:r_i_r_f}. Solving Eq.~\eqref{eq:r_i_r_f} with shell crossing is complicated. Instead, we keep Eq.~\eqref{eq:r_i_r_f} unchanged, but modify $\Delta \epsilon$ to avoid shell crossing. Note that this does not mean shell crossing does not happen, but each mass shell is now interpreted an an effective mass shell after the new equilibrium is reached. We first compute the ratio
\begin{equation}
\xi(r_\mathrm{i}) = \frac{\Delta \epsilon(r_\mathrm{i})}{ \mathrm{G} M_\mathrm{i} / 2 r_\mathrm{i}}.
\end{equation}
To avoid $\mathrm{d}r_\mathrm{f}/\mathrm{d}r_\mathrm{i} < 0$, we assume that at small radii, i.e. $r < r_{\rm crossing}$, $\xi(r_\mathrm{i})$ is constant such that $\Delta \epsilon(r_\mathrm{i})$ is proportional to the the gravitational potential:
\begin{equation}
\Delta \epsilon(r_\mathrm{i}) = \xi(r_{\rm crossing}) \frac{G M_\mathrm{i}}{2 r_\mathrm{i}}.
\label{eq:heating_mod}
\end{equation}
To ensure that, after tidal heating, the density profile remains continuous at $r_{\rm crossing}$,~\footnote{From Eqs.~\eqref{eq:r_i_r_f} and \eqref{eq:rho_f}, this requires that $\Delta \epsilon$ is differentiable.} we require that both $\xi$ and $\mathrm{d}\xi/\mathrm{d}r$ are continuous at $r_{\rm crossing}$ or equivalently
\begin{eqnarray}
\xi(r_{\rm crossing}) &=& \frac{\Delta \epsilon(r_{\rm crossing})}{ \mathrm{G} M(r< r_{\rm crossing}) / 2 r_{\rm crossing}},
\label{eq:r_cross_1}\\
\left.\frac{\mathrm{d} \xi(r_\mathrm{i})}{\mathrm{d}r_\mathrm{i}}\right|_{r_\mathrm{i}=r_{\rm crossing}} &=& 0.
\label{eq:r_cross_2}
\end{eqnarray}
The shell crossing radius, $r_{\rm crossing}$, and $\xi(r_{\rm crossing})$ are uniquely determined by Eqs.~\eqref{eq:r_cross_1} and \eqref{eq:r_cross_2}.

\begin{figure}
\includegraphics[width=\columnwidth]{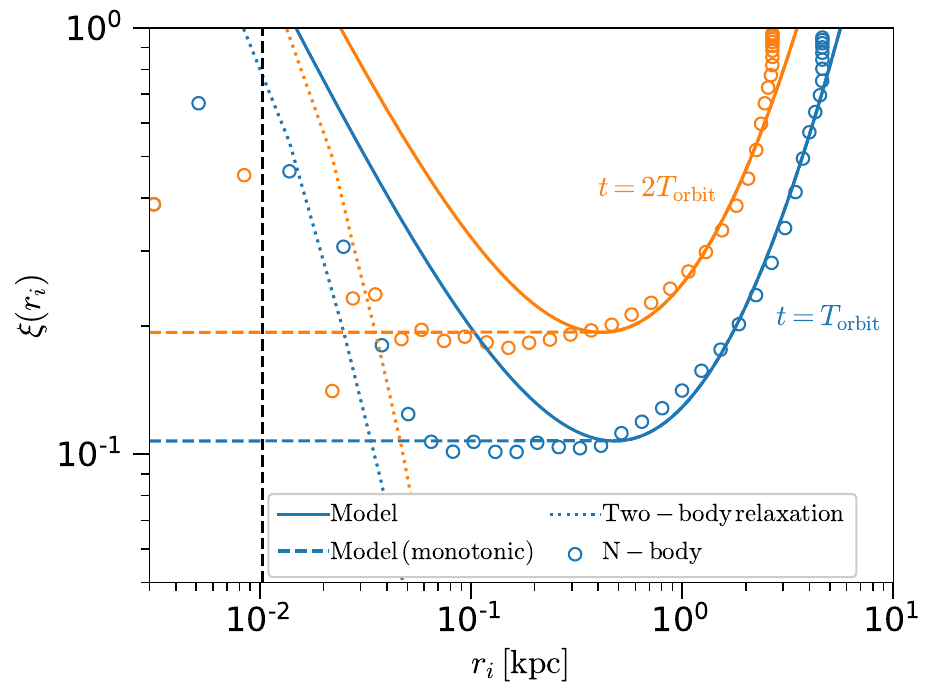}
\caption{Heating energy ratio, $\xi$, from our model Eq.~\eqref{eq:heating_improve} (solid curve) and that includes the monotonic correction Eq.~\eqref{eq:heating_mod} (dashed curves), compared with that measured directly from simulations (colored circles). The colored dotted curves show the heating energy from two-body relaxation.}
\label{fig:E_heating}
\end{figure}

Figure~\ref{fig:E_heating} shows the ratio $\xi$ as a function of the initial radius of mass shells computed from Eqs.~\eqref{eq:heating_improve} and \eqref{eq:heating_mod} compared with that measured from simulations. Here, we consider a cored subhalo with $(\alpha,\beta,\gamma)=(1,3,0)$ and $R_p/R_a=2/5$. As can be seen, the model prediction without the monotonic correction (solid curves) overestimates the heating energy ratio at small radii. On the other hand, the heating energy measured from simulations (colored circles) flattens out with decreasing radius, which verifies our assumption Eq.~\eqref{eq:heating_mod} (see the dashed curves). At radii close to the softening length (vertical line), the effects of two-body relaxation~\cite{2008gady.book.....B} (dotted curves) become significant leading to the rise of $\xi$. Here, we compute the heating specific energy due to two-body relaxation as
\begin{equation}
\Delta \epsilon_{\mathrm{two-body}}(r)=2\alpha_{\mathrm{two-body}}\frac{\ln\Lambda}{N(r)} \frac{v(r)^3}{r}(t-t_0),
\label{eq:two_body}
\end{equation}
where $\alpha_{\mathrm{two-body}}$ is the efficiency of two-body relaxation heating, $v(r)$ is the circular velocity, $t$ and $t_0$ are the current and initial time, respectively, $N(r)$ is the number of particles within radius $r$, and the Coulomb logarithm is computed as
\begin{equation}
\ln\Lambda = \frac{1}{2}\ln\left[1+\left(\frac{r}{\max\left\{r_{\mathrm{softening}},\sqrt{\frac{2 G M_\mathrm{p}}{v^2}}\right\}}\right)^2\right].
\label{eq:coulomb_two_body}
\end{equation}
To compute the contribution of two-body relaxation to the heating energy in Fig.~\ref{fig:E_heating}, we take $\alpha_{\mathrm{two-body}}=0.006$. 
We have also checked the case with $\gamma=0.5$, for which similar behavior in $\xi$ is found. For $\gamma\geq 1$, $\xi(r_i)$ is monotonic, thus no correction is needed, but the two-body relaxation heating is also dominated at very small radii, leading to the formation of an artificial core in halo center.

\subsection{Calibrations}\label{subsec:calibration}

\begin{table*}
\begin{center}
\caption{Priors and median values (with $16$th and $84$th percentiles) for model parameters (rows) for dark matter subhalos with different inner slopes (columns).}
\label{tab:best_fit_mcmc}
\begin{tabular}{c|c|r@{.}l|r@{.}l|r@{.}l|r@{.}l}
\hline
             & \textbf{Prior}   & \multicolumn{2}{c}{\boldmath{$\gamma=0$}} & \multicolumn{2}{c}{\boldmath{$\gamma=0.5$}} & \multicolumn{2}{c}{\boldmath{$\gamma=1$}
             } & \multicolumn{2}{c}{\boldmath{$\gamma=1.5$}
             } \\
\hline
$\alpha_\mathrm{s}$      & $[0,20]$  & $4$&$4^{+8.5}_{-2.3}$      & $1$&$06^{+0.11}_{-0.089}$  & $3$&$93^{+0.77}_{-0.58}$       & $>16$&$2$                      \\
$\epsilon_\mathrm{h}$    & $[0,6]$   & $0$&$262^{+0.042}_{-0.028}$ & $0$&$166^{+0.013}_{-0.012}$  & $0$&$0741^{+0.0052}_{-0.0047}$ & $0$&$0403^{+0.0093}_{-0.0084}$ \\
$f_2$                    & $[0,2]$   & $0$&$21^{+0.039}_{-0.035}$  & $0$&$166^{+0.023}_{-0.023}$  & $0$&$547^{+0.043}_{-0.041}$    & $1$&$04^{+0.22}_{-0.18}$       \\
$\beta_\mathrm{h}$       & $[0,4]$   & $0$&$37^{+0.16}_{-0.11}$    & $0$&$0986^{+0.047}_{-0.044}$ & $0$&$278^{+0.039}_{-0.034}$ & $0$&$358^{+0.054}_{-0.047}$       \\
$\ln\mathcal{R}$         & $[-14,0]$ & $-2$&$13^{+0.12}_{-0.11}$   & $-1$&$92^{+0.11}_{-0.096}$   & $-2$&$77^{+0.085}_{-0.08}$     & $2$&$52^{+0.08}_{-0.077}$      \\
\hline
\end{tabular}
\end{center}
\end{table*}

In the semi-analytic models presented in the previous subsection, there are in total four free parameters that must be calibrated to simulations. One of these parameters is $\alpha_\mathrm{s}$ in the tidal stripping model, the others are the parameters $\epsilon_\mathrm{h}$, $f_2$, and $\beta_\mathrm{h}$ in the tidal heating model. We calibrate these parameters by comparing the predictions for $M_{\rm bound}$, $V_{\mathrm{max}}$ and $R_{\mathrm{max}}$ from our models to the results measured from simulations presented in Sec.~\ref{sec:results}. To perform this calibration we define a likelihood function as
\begin{eqnarray}
&&\ln \mathcal{L}(M_{\rm bound},V_{\rm max},R_{\rm max}|\alpha,\epsilon_\mathrm{h},f_2,\beta_\mathrm{h}) \nonumber\\
=&& -\frac{1}{2}\sum_i\left[\left(\frac{M_{\rm bound, i}^\mathrm{p}-M_{\rm bound, i}}{\sigma_{M,i}}\right)^2+ \ln\left(2\pi \sigma_{M,i}^2\right) \right]\nonumber \\
&&-\frac{1}{2}\sum_i\left[\left(\frac{V_{\rm max, i}^\mathrm{p}-V_{\rm max, i}}{\sigma_{V,i}}\right)^2 + \ln\left(2\pi \sigma_{\rm V, i}^2\right) \right]\nonumber \\
&&-\frac{1}{2}\sum_i\left[\left(\frac{R_{\rm max, i}^\mathrm{p}-R_{\rm max, i}}{\sigma_{R,i}}\right)^2 + \ln\left(2\pi \sigma_{R, i}^2\right) \right],
\label{eq:logL}
\end{eqnarray}
where $M_{\rm bound, i}^\mathrm{p}$, $V_{\rm max, i}^\mathrm{p}$, and $R_{\rm max, i}^\mathrm{p}$ are model predictions at the i$^\mathrm{th}$ snapshot. Here, $\sigma_{M,\mathrm{i}}$, $\sigma_{V,\mathrm{i}}$ and $\sigma_{R,\mathrm{i}}$ represent the combined uncertainties in the measurements and the model. Given that our models, like any models, are imperfect, we introduce a free parameter $\mathcal{R}$ that quantifies the model uncertainties and write the total uncertainties as
\begin{eqnarray}
\sigma_{M,\mathrm{i}}^2&=&\widetilde{\sigma}_{M,\mathrm{i}}^2+(\mathcal{R} M_{\rm bound,\mathrm{i}}^\mathrm{p})^2,\\
\sigma_{V,\mathrm{i}}^2&=&\widetilde{\sigma}_{V,\mathrm{i}}^2+(\mathcal{R} V_{\rm max,\mathrm{i}}^\mathrm{p})^2,\\
\sigma_{R,\mathrm{i}}^2&=&\widetilde{\sigma}_{R,\mathrm{i}}^2+(\mathcal{R} R_{\rm max,\mathrm{i}}^\mathrm{p})^2.
\end{eqnarray}
Here $\widetilde{\sigma}_{M,\mathrm{i}}$ and $\widetilde{\sigma}_{V,\mathrm{i}}$ are the Poisson errors measured from simulations, and $\widetilde{\sigma}_{R,\mathrm{i}}$ is defined as half of the radial bin width used for computing $V_{\mathrm{max}}$.~\footnote{Note that we have performed a supersampling of the subhalo profiles, thus the bin width used here is smaller than the original radial bin width (see Sec.~\ref{subsec:analysis}).}

We run Monte Carlo Markov Chain (MCMC) simulations for dark matter profiles with different density slopes at small radii, from cored profiles ($\gamma=0$) to very cuspy profiles ($\gamma=1.5$). We refrain from performing MCMC simulations for all the combinations of $(\alpha,\beta,\gamma)$ shown in Sec.~\ref{sec:results} as we find that the model parameters are mostly sensitive to the inner slope of dark matter halo. We have checked that our models also work well for other choices of the parameter that control the outer profile of the subhalos. In the remainder of this section, we set $(\alpha,\beta)=(1,3)$.

\begin{figure*}
\includegraphics[width=1.8\columnwidth]{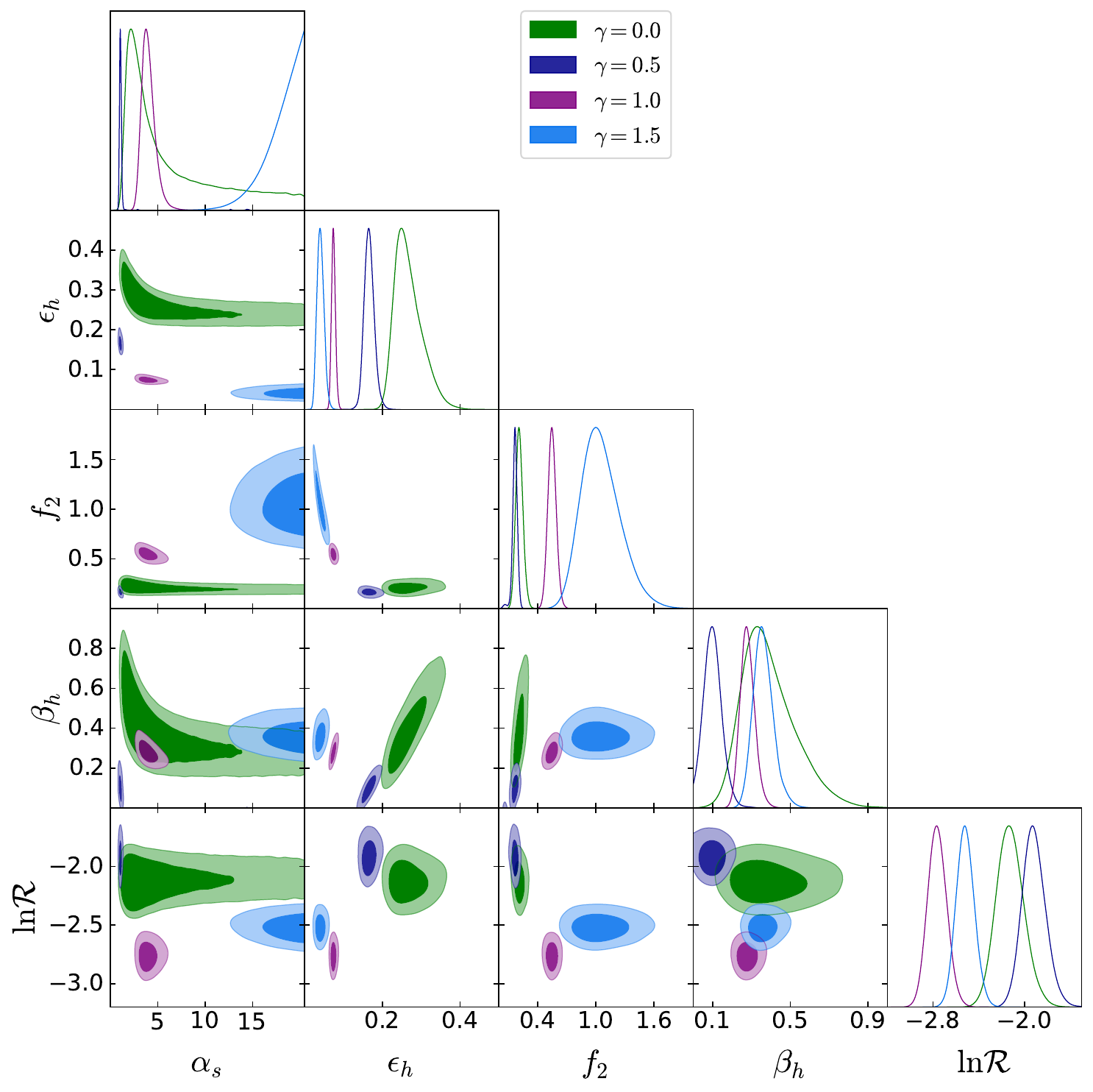}
\caption{Posterior distributions over model parameters for different values of $\gamma$.}
\label{fig:posterior}
\end{figure*}

\begin{figure}
\includegraphics[width=\columnwidth]{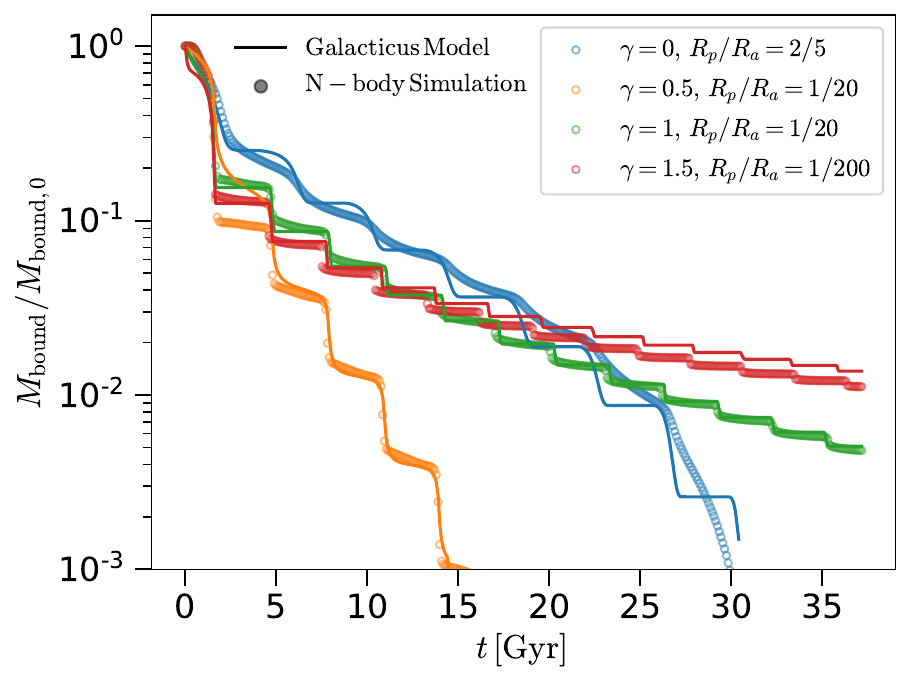}
\caption{Bound mass as a function of time from semi-analytic models (lines) compared with $N$-body simulations (circles). Colors indicate different values of the inner slope of the density profile, $\gamma$, as indicated in the figure. Other parameters in the density profile are fixed at $(\alpha,\beta)=(1,3)$.}
\label{fig:Mbound}
\end{figure}

For model parameters $\{\alpha_s,\epsilon_h,f_2,\beta_h\}$ we adopt uniform priors. For $\mathcal{R}$, a loguniform prior is used. The priors and resulting median values are listed in Table~\ref{tab:best_fit_mcmc}. The posteriors of model parameters for different dark matter profiles are shown in Fig.~\ref{fig:posterior}. We find that for $\gamma=1.5$, the tidal stripping efficiency parameter $\alpha_s$ is unconstrained from above. We report the 16th percentile in Table~\ref{tab:best_fit_mcmc} as a conservative lower bound. In practice, this means that our model might have underestimated the tidal radius $r_t$ in this case such that there is not enough mass outside $r_t$ to be stripped. Including partial of the contribution from the centrifugal force in computing $r_t$, i.e.\ assuming a nonzero value of $\gamma_s$ in Eq.~\eqref{eq:r_t}, may help improve the fitting. Nevertheless, our current model already fits the bound mass evolution very well for $\gamma=1.5$, see Fig.~\ref{fig:Mbound}.

\begin{figure}
\includegraphics[width=\columnwidth]{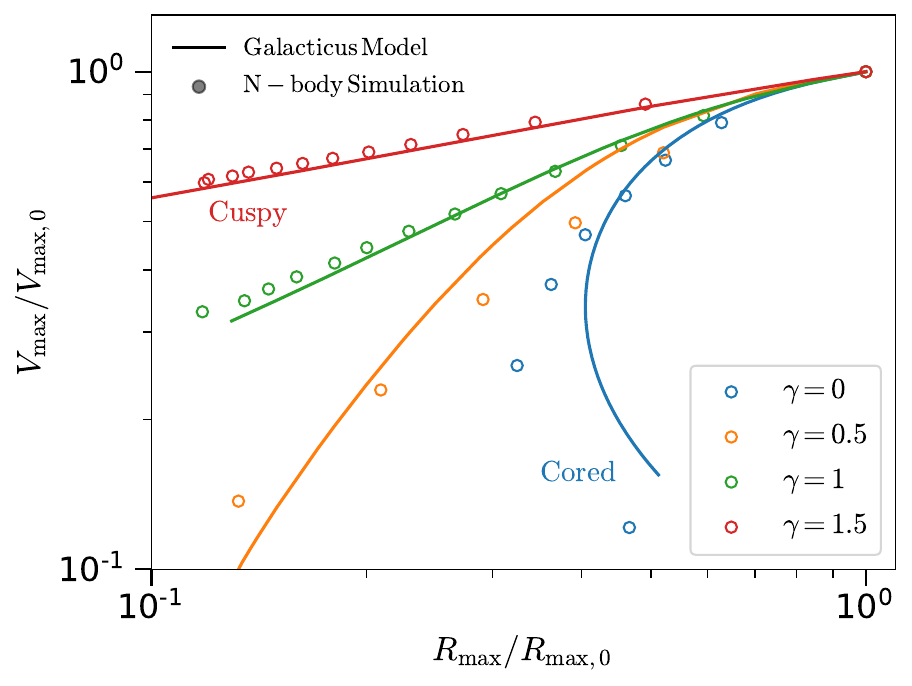}
\caption{Tidal tracks from semi-analytic models (lines) compared with $N$-body simulations (circles). Colors indicate different values of the inner slope of the density profile, $\gamma$, as indicated in the figure.}
\label{fig:tidal_track_compare}
\end{figure}

In Fig.~\ref{fig:tidal_track_compare}, we show the predicted tidal tracks for different $\gamma$ from our best-fit models compared with the results from $N$-body simulations. For the cases with $\gamma=1$ and $\gamma=1.5$, our models agree very well with the simulation results. For the cases with $\gamma=0.5$ and $\gamma=0$, the agreement is somewhat worse, but nevertheless captures the overall behavior reasonably well. Notably, for the cored case ($\gamma=0$), our model reproduces the turnaround of the tidal track when the subhalo is heavily stripped.

\begin{figure*}
\includegraphics[width=\columnwidth]{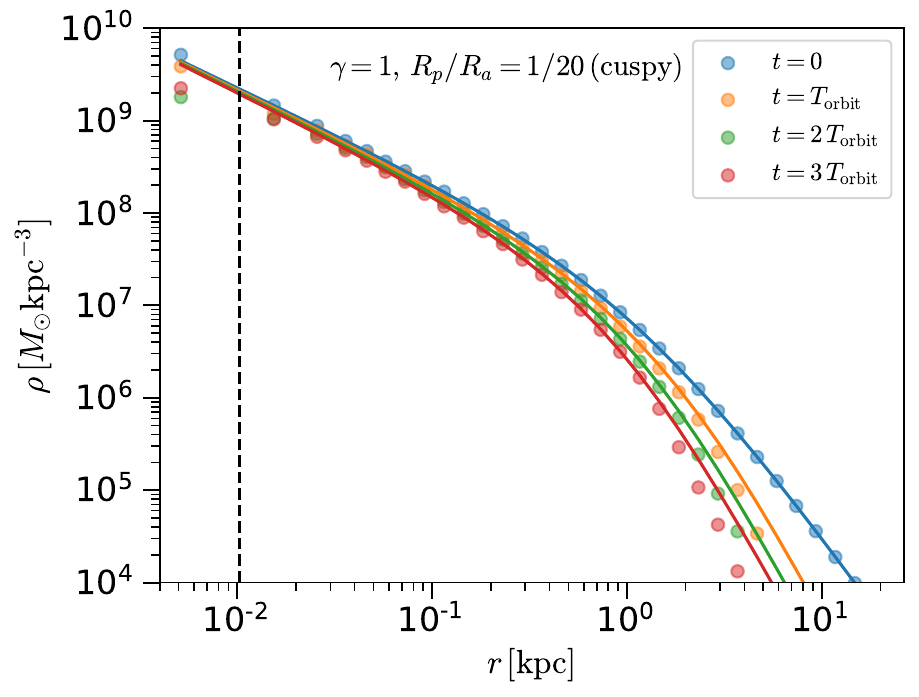}
\includegraphics[width=\columnwidth]{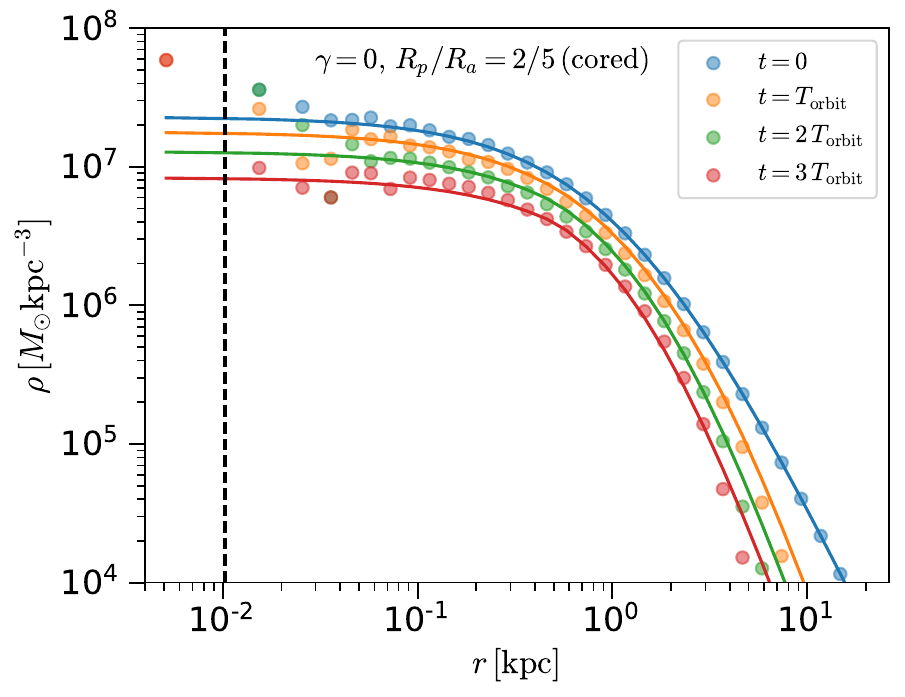}
\caption{Density evolution of cuspy (left panel) and cored subhalos (right panel) from semi-analytic models (lines) compared with $N$-body simulations (circles). Colors indicate different times.}
\label{fig:densisty_sams}
\end{figure*}

In Fig.~\ref{fig:Mbound}, we also show the bound mass evolution for different $\gamma$ from our best-fit models compared with the results from $N$-body simulations. The corresponding density profiles at different times for $\gamma=1$ and $\gamma=0$ are shown in Fig.~\ref{fig:densisty_sams}.

Ideally, we would expect that the values of the model parameters should be consistent across different dark matter profiles, if our model correctly captures the dependence of tidal stripping and heating on the subhalo density profile. However, we find that there does not exist a single set of model parameters that fit all the cases accurately (see Fig.~\ref{fig:posterior}). Thus we report the best-fit parameters for each dark matter profile separately. This also suggests that there are additional dependencies on the inner slope of dark matter halo profiles that are not fully captured by our current model. For $\gamma$ that is not listed in Table~\ref{tab:best_fit_mcmc}, we suggest doing an interpolation. We defer exploration of a universal model to future work.

\section{Conclusions and discussions}\label{sec:cons}

We have run high-resolution idealized simulations to study the evolution of dark matter subhalos under the tidal effects from their host. We consider a generalized dark matter halo profile controlled by three parameters $\alpha$, $\beta$, and $\gamma$ [see Eq.~\eqref{eq:prof_abg}]. By changing these parameters, we can represent a dark matter profile with a flat core ($\gamma=0$) or NFW-like cuspy profile ($\gamma=1$). We have run simulations with different combinations of these parameters and found the fitting functions for the tidal track in each case. The $V_{\rm max}-R_{\rm max}$ tracks we find for different $\gamma$, the inner slope of dark matter profile, are in  agreement with the previous studies by Ref.~\cite{Penarrubia:2010jk} expect for the case with a extremely cuspy profile, i.e. $\gamma=1.5$. We have checked the convergence of the results for $\gamma=1.5$ and found that using a global time step size leads to better converged results. Our converged results still show some differences from the one in Ref.~\cite{Penarrubia:2010jk}. We note that Ref.~\cite{Penarrubia:2010jk} uses a global, fixed time step size, similar to the global time step size in our convergence tests. We find that their time step size may be too large to obtain converged tidal track.

We have also run tests with lower subhalo concentration and a host potential that grows with time, see Appendix~\ref{appendix:diff_c}. While the bound mass of subhalos in these cases evolves differently than the fiducial settings (i.e. lower concentration subhalos are more influenced by the tidal effects), the tidal tracks are only marginally affected, confirming that the tidal tracks are mostly sensitive to the bound mass fraction. Furthermore, we find that adding a Miyamoto and Nagai disk~\cite{Miyamoto:1975} and Hernquist bulge~\cite{Hernquist:1990be} potential to the host to mimic the Milky Way disk and bugle also does not have a significant impact on the tidal track---the difference from the fiducial case is less than $10\%$ (see Appendix~\ref{appendix:disk}).

From the simulation data, we measure the density transfer function that connects the current density profile of a subhalo to its initial profile. Similar to previous studies~\cite{Hayashi:2002qv,Penarrubia:2007zx,Penarrubia:2010jk,Green:2019zkz,Errani:2020wgn}, we find that the transfer function is mainly sensitive to the bound mass fraction and is insensitive to the subhalo orbit. Using a similar fitting formula to that proposed in Ref.~\cite{Hayashi:2002qv} for the transfer function, we find the effective tidal radius, $r_{\mathrm{te}}$, and normalization parameter, $f_\mathrm{t}$, (see Eq.~\eqref{eq:transfer}) and give  fitting formula for $r_{\mathrm{te}}$ and $f_\mathrm{t}$ as functions of bound mass fraction, see Eqs.~\eqref{eq:rte_new} and \eqref{eq:ft_new}. These transfer function fits can be used to model the density evolution of subhalo with a variety of density profiles.

We then present improved semi-analytic models for tidal stripping and tidal heating built on our previous work \cite{2022MNRAS.517.1398B}. In our previous work~\cite{2022MNRAS.517.1398B}, the semi-analytic models were calibrated only to dark matter halos with NFW profiles. In this work, we extend the calibration to other profiles. We find that the no shell crossing assumption in our previous tidal heating model is not valid for cored dark matter profiles. To overcome this issue, we propose a simple modification to the heating energy. The modified model is shown to work well for cored dark matter profiles.

For CDM, it is well known that the dark matter halos are well described by the NFW profile. But for other types of dark matter particle this is not necessarily true. For example, for SIDM, due to frequently scattering between dark matter particles in the halo center, a constant density core can form. At later stages of the evolution of SIDM halos, core collapse can happen leading to a very cuspy density profile. Core formation can also happen in other dark matter models such as fuzzy dark matter due to additional pressure from the quantum effects or, in CDM models, due to baryonic feedback. Thus considering the evolution for different dark matter profiles is useful to allow comparison of different dark matter models with observations of subhalos. Using the semi-analytic models presented in this work will allow us to predict the statistical properties of subhalos in different dark matter models and distinguish these models by comparing with observations such as the Milky Way satellite populations and strong gravitational lenses.

One limitation of the current work is that we consider only one typical host mass of $10^{12} \mathrm{M}_{\odot}$ with NFW profile and a fixed subhalo mass of $10^9 \mathrm{M}_{\odot}$. The concentration of the host and subhalo are also fixed. Although the tidal tracks are not very sensitive to changes in the host properties, they have a weak dependence on the concentration of subhalos~\cite{Green:2019zkz}. Thus the semi-analytic models presented in this work need to be tested against a larger set of simulations covering a range of halo masses and concentrations. It will also be useful to test the calibrated models against cosmological simulations as done in~\cite{Yang:2022zkd,2020MNRAS.498.3902Y} and take into account the pre-infall tidal effects~\cite{Behroozi:2013fqa,Wetzel:2015xia,Nadler:2020ulu}. A more detailed study on this will be presented in a forthcoming paper.

Furthermore, in the current work, we have ignored non-gravitational interactions between dark matter particles. In future works, we will explore the possibility to include other effects in different dark matter scenarios, e.g. enhanced tidal stripping in fuzzy dark matter models due to ``quantum tunneling"~\cite{Hui:2016ltb,Du:2018qor}, core evolution~\cite{Balberg:2002ue,Ahn:2004xt,Koda:2011yb,Yang:2022zkd,Outmezguine:2022bhq} and ram pressure stripping~\cite{Kummer:2017bhr} in SIDM model.

\section*{Acknowledgements}
X.D. thanks Jorge Pe\~{n}arrubia for beneficial discussions on initial conditions and simulation with {\sc SUPERBOX}.
X.D. and T.T. acknowledge support from the National Science Foundation through Grant No. NSF-AST-1836016 and NSF-AST-2205100, and by the Gordon and Betty Moore Foundation through Grant No. 8548. A. N. acknowledges support from the National Science Foundation through Grant No. NSF-AST-2206315.

Computing resources used in this work were made available by a generous grant from the Ahmanson Foundation.


\appendix
\section{Evolution of $R_{\mathrm{max}}$ and $V_{\mathrm{max}}$ as functions of the bound mass fraction}\label{appendix:track_more}

\begin{figure*}
\includegraphics[width=0.9\columnwidth]{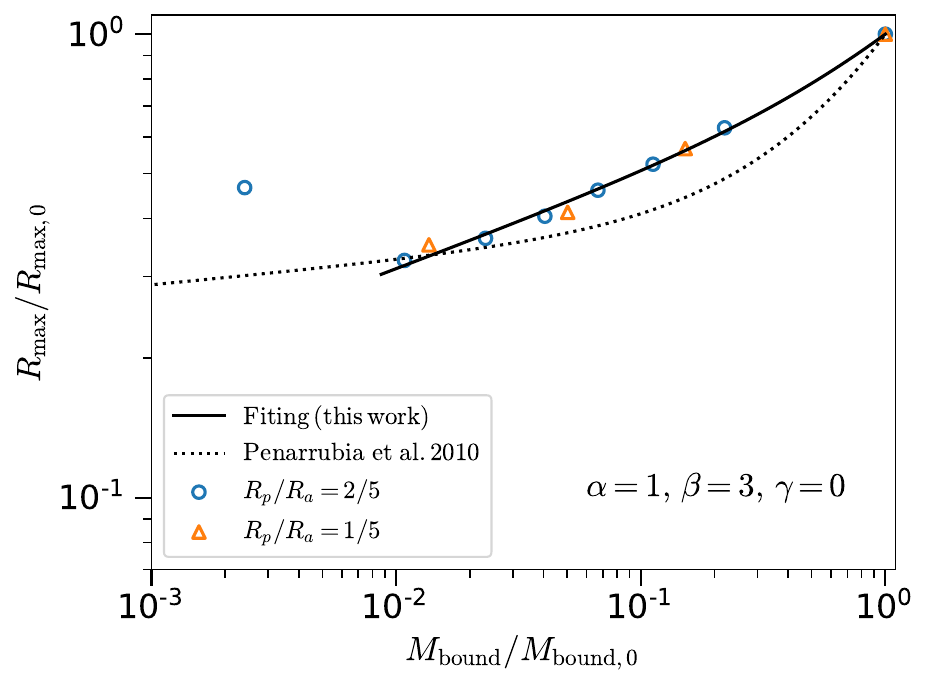}
\includegraphics[width=0.9\columnwidth]{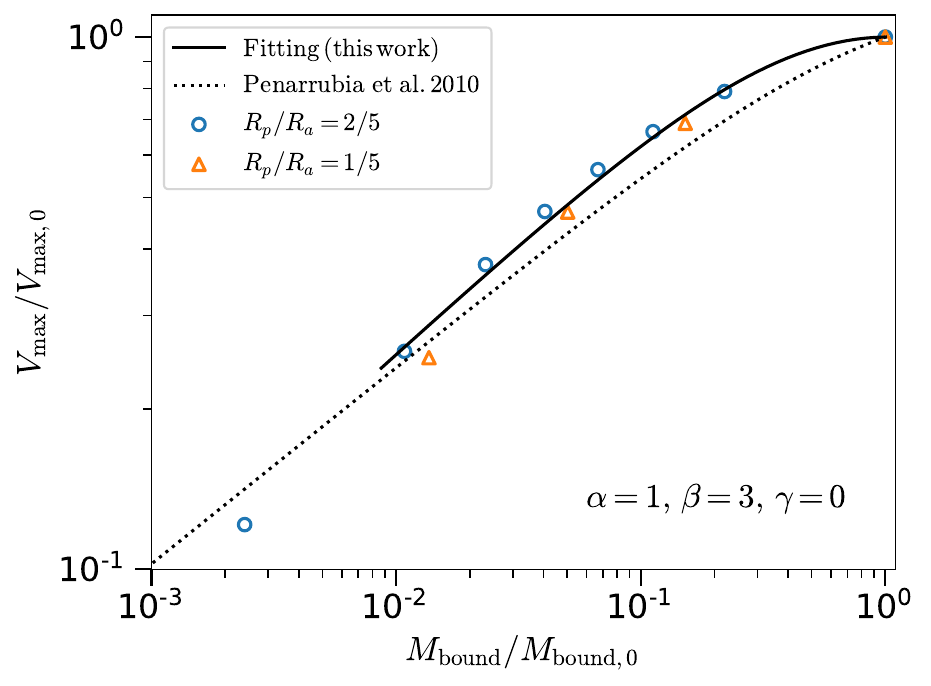}
\includegraphics[width=0.9\columnwidth]{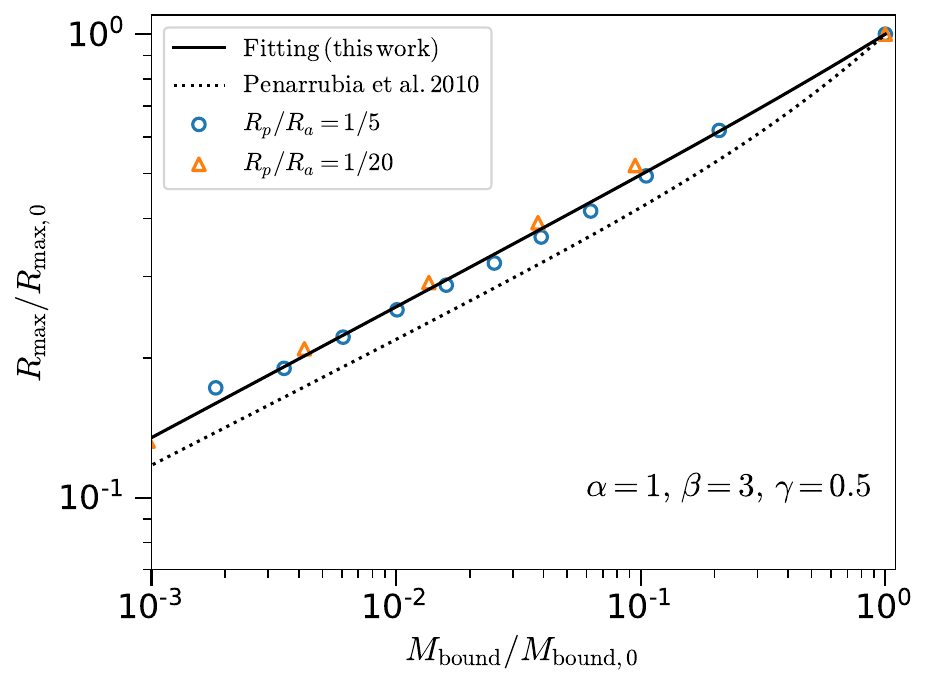}
\includegraphics[width=0.9\columnwidth]{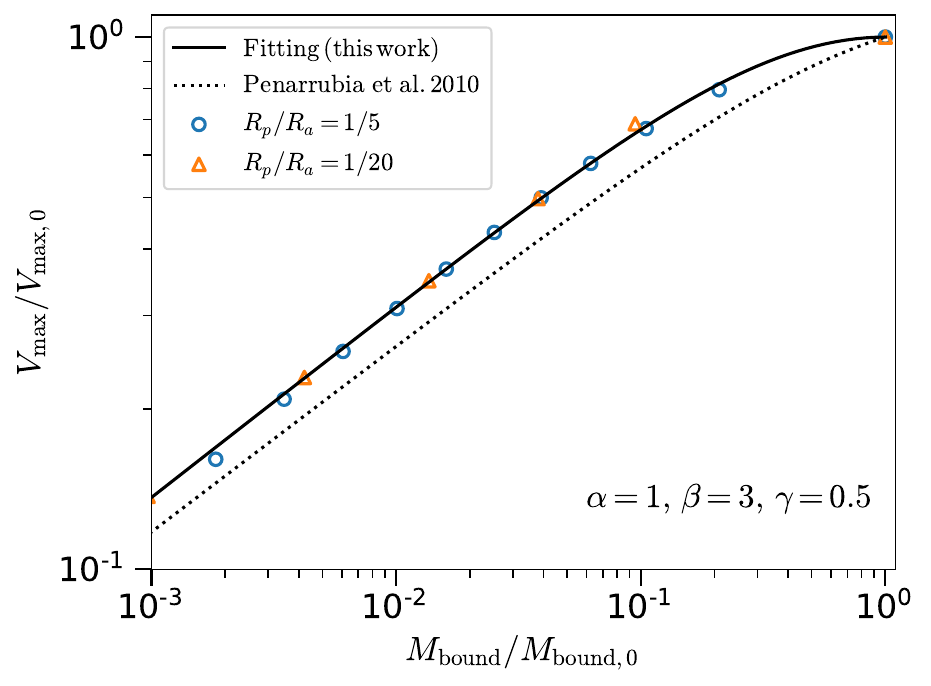}
\includegraphics[width=0.9\columnwidth]{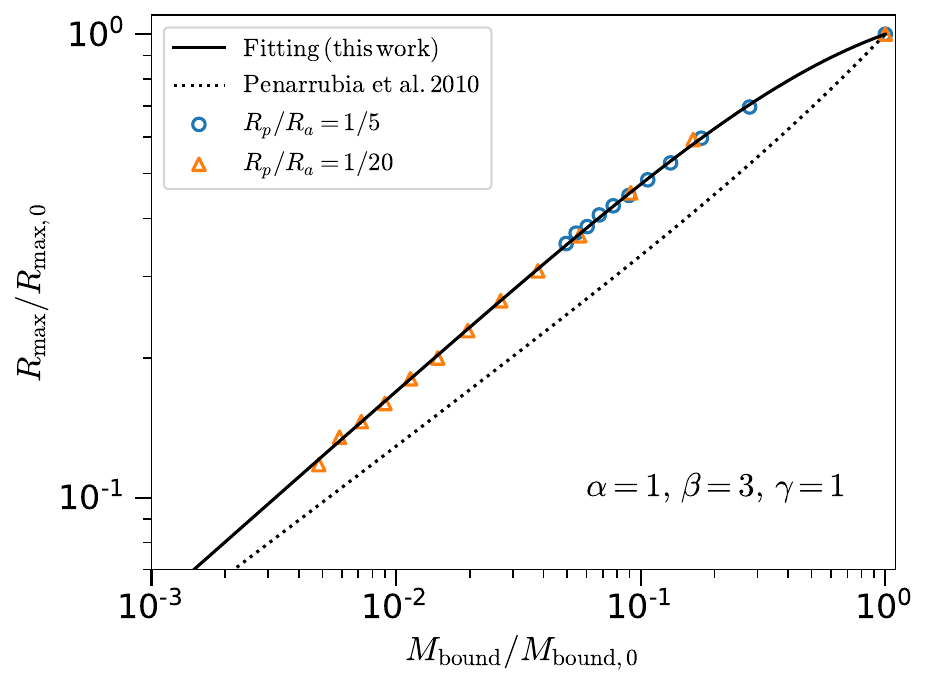}
\includegraphics[width=0.9\columnwidth]{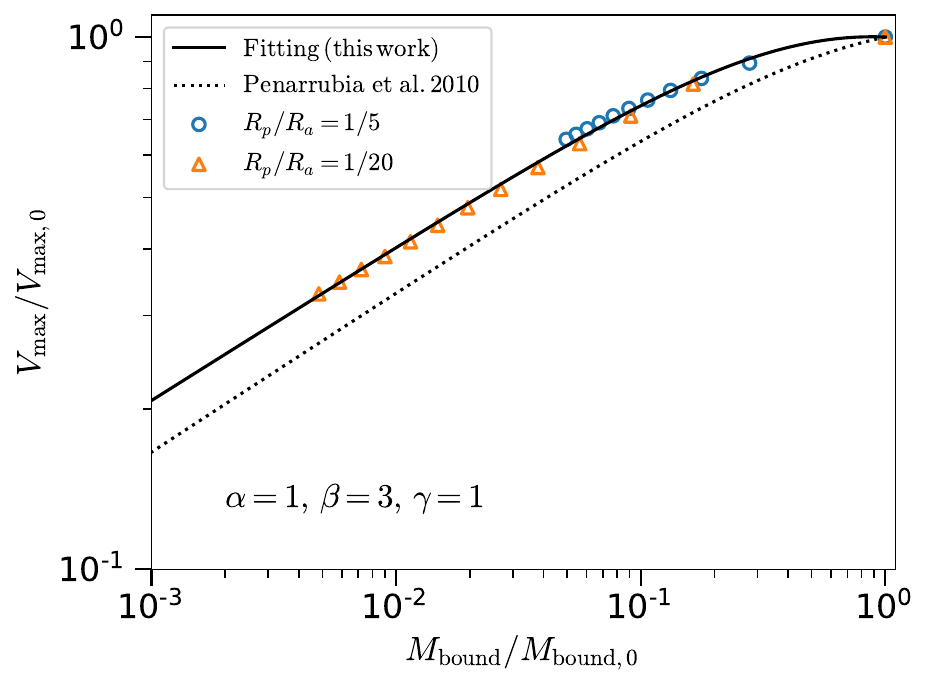}
\includegraphics[width=0.9\columnwidth]{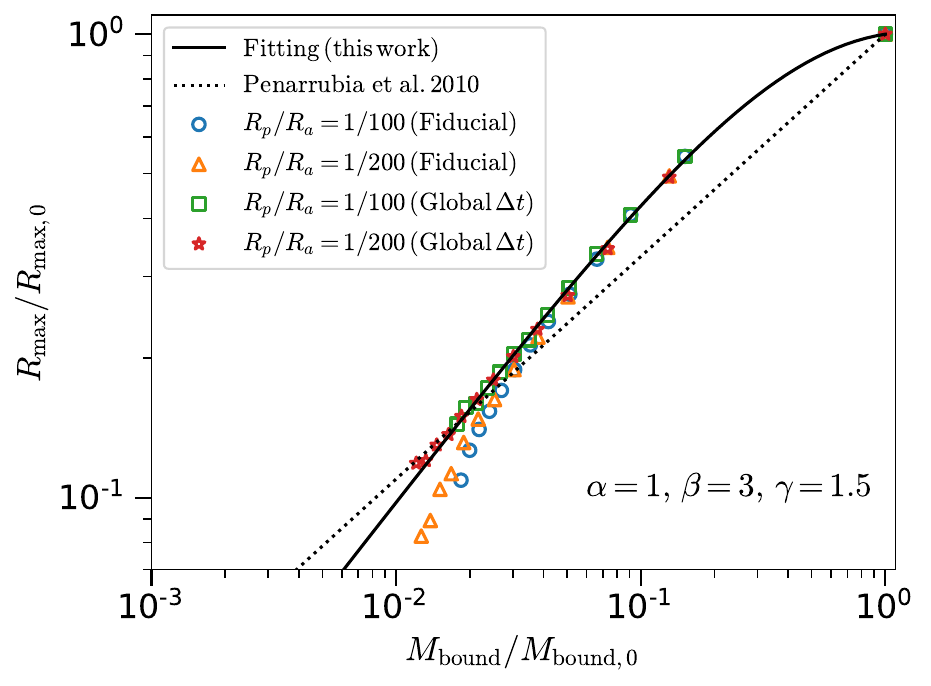}
\includegraphics[width=0.9\columnwidth]{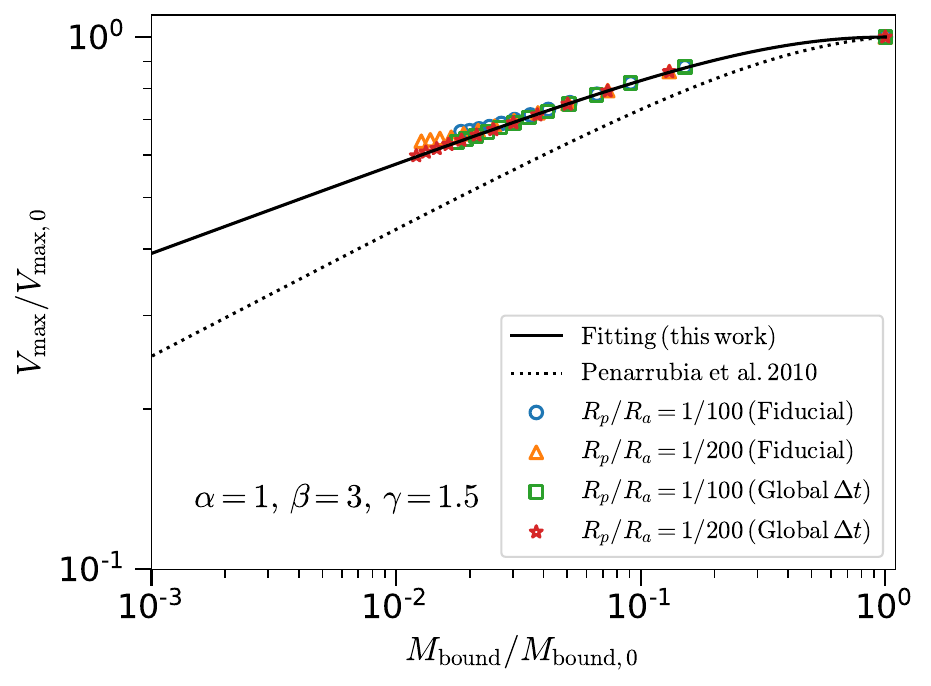}
\caption{$R_{\mathrm{max}}$ and $V_{\mathrm{max}}$ for $(\alpha,\beta)=(1,3)$ and different values of $\gamma$ from $N$-body simulations using {\sc Gadget-4}. Solid lines show the fitting functions found in this work.}
\label{fig:Rmax_Vmax_M_1}
\end{figure*}

\begin{figure*}
\includegraphics[width=0.9\columnwidth]{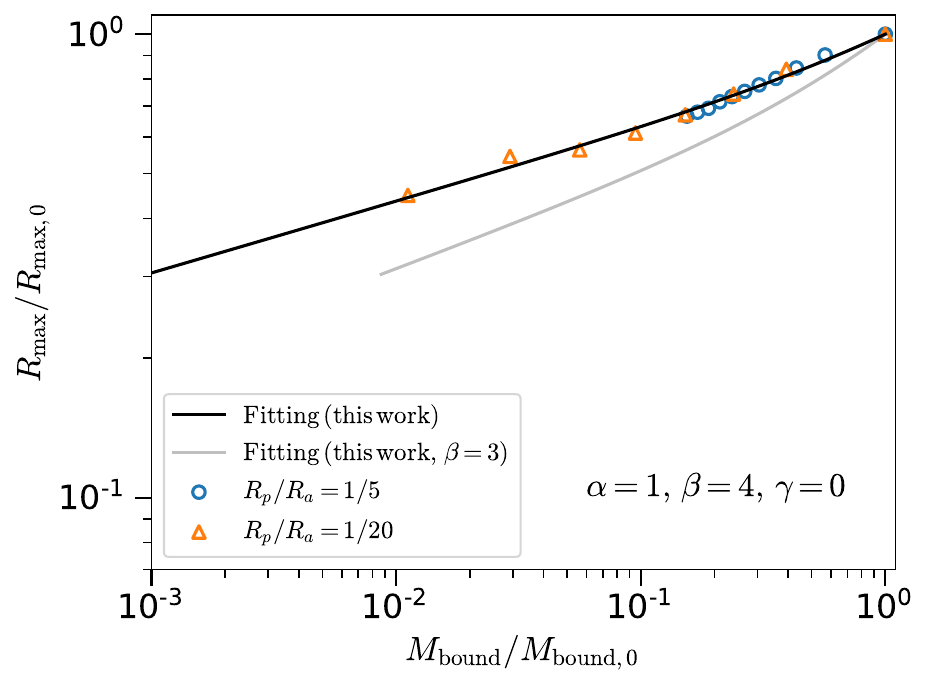}
\includegraphics[width=0.9\columnwidth]{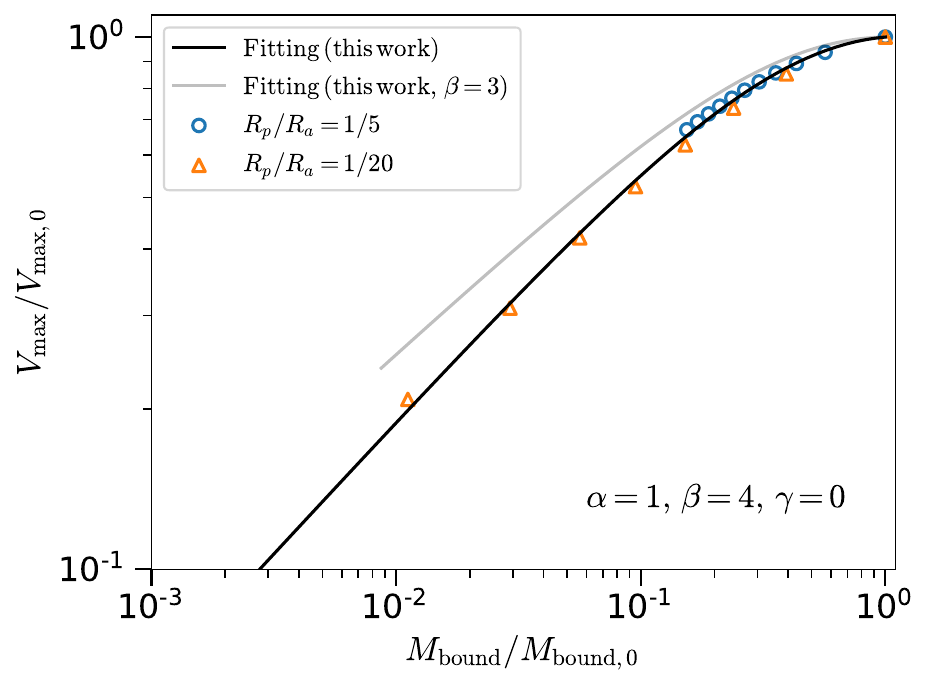}
\includegraphics[width=0.9\columnwidth]{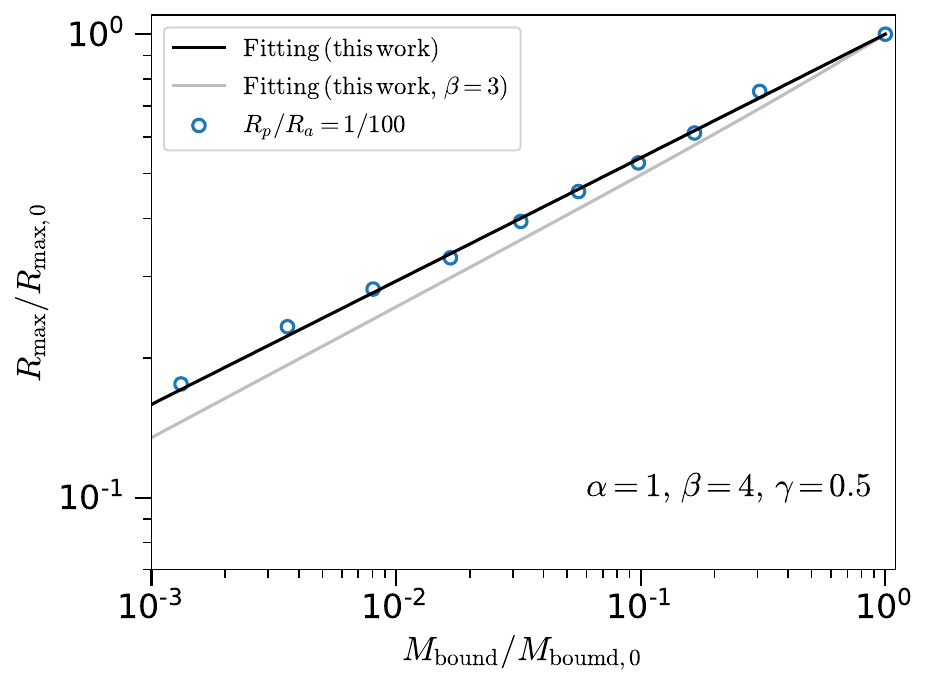}
\includegraphics[width=0.9\columnwidth]{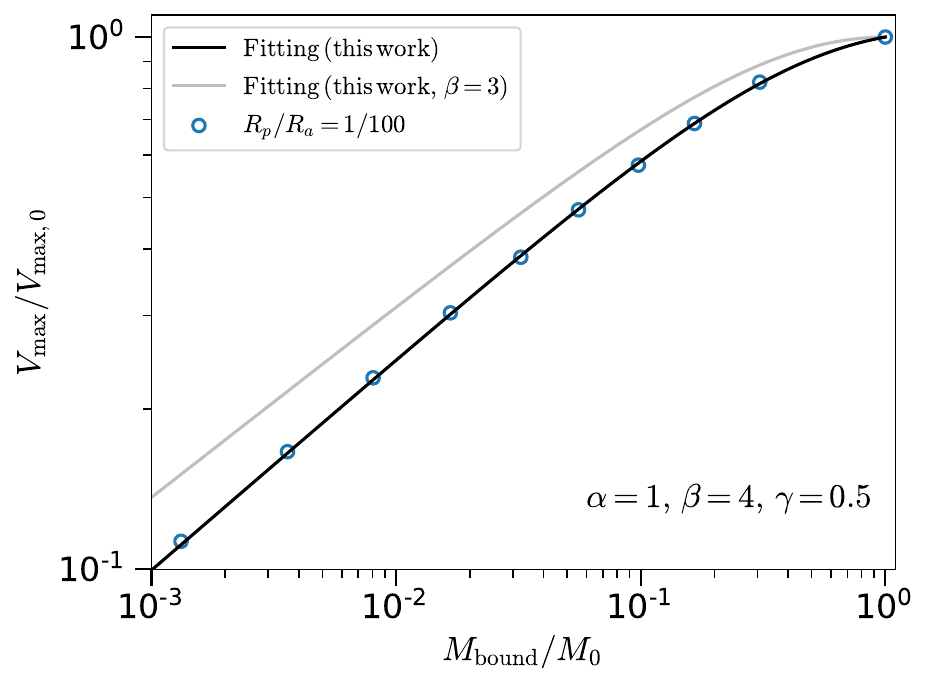}
\includegraphics[width=0.9\columnwidth]{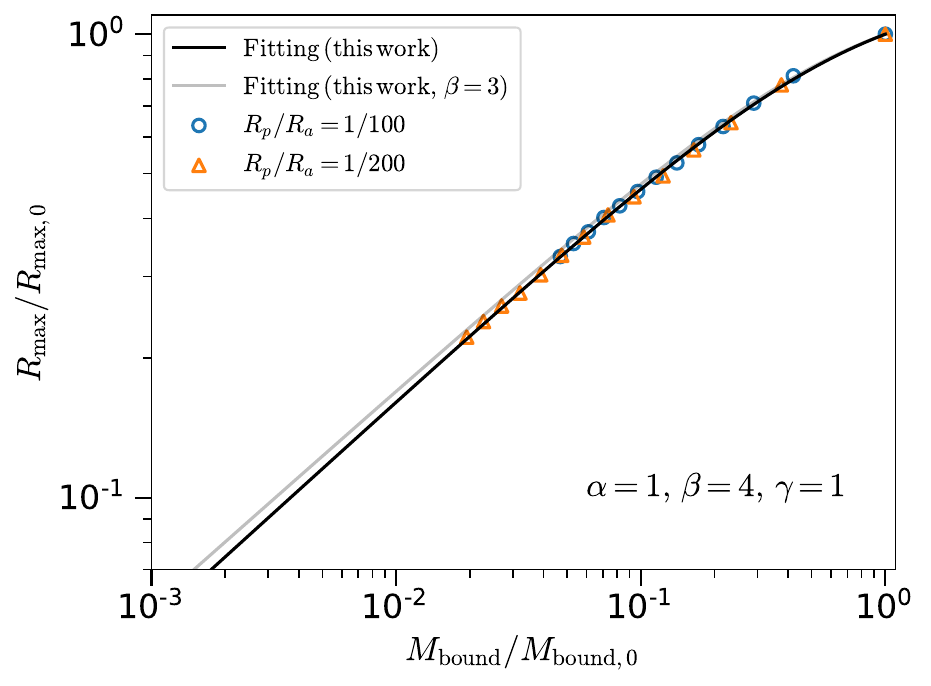}
\includegraphics[width=0.9\columnwidth]{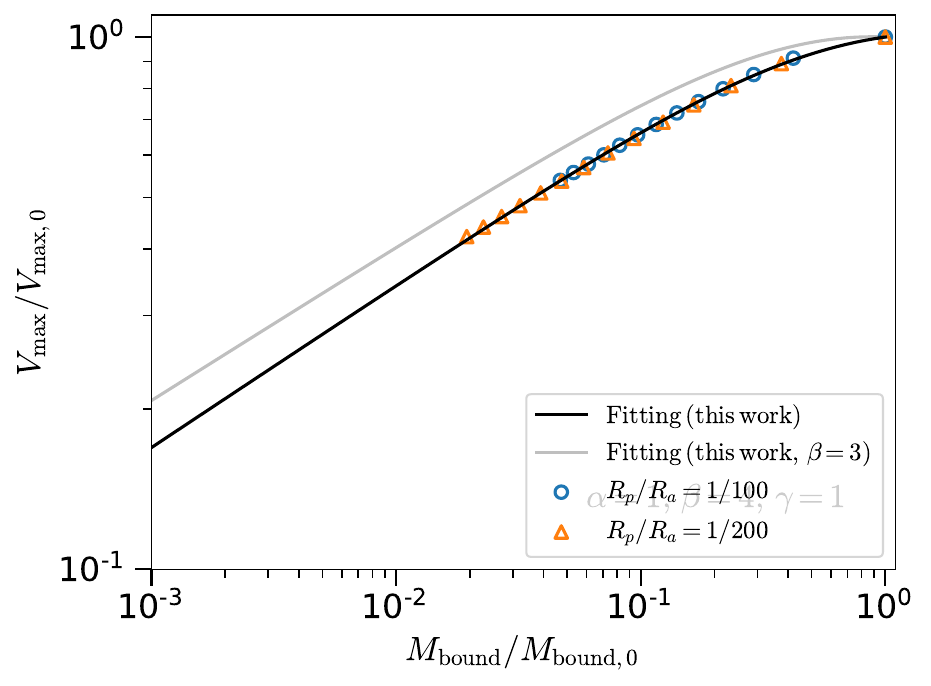}
\includegraphics[width=0.9\columnwidth]{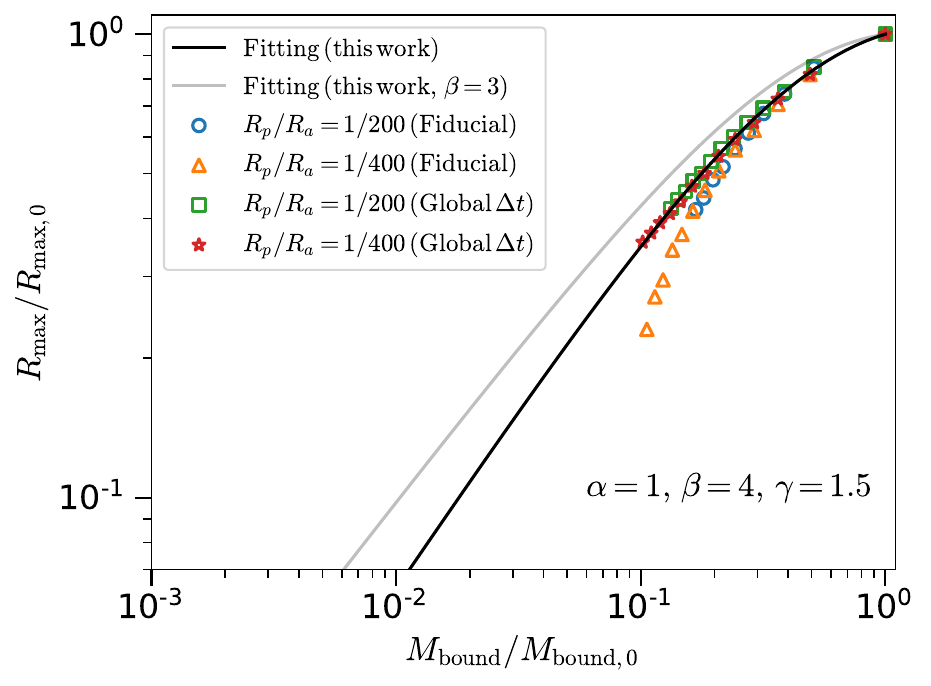}
\includegraphics[width=0.9\columnwidth]{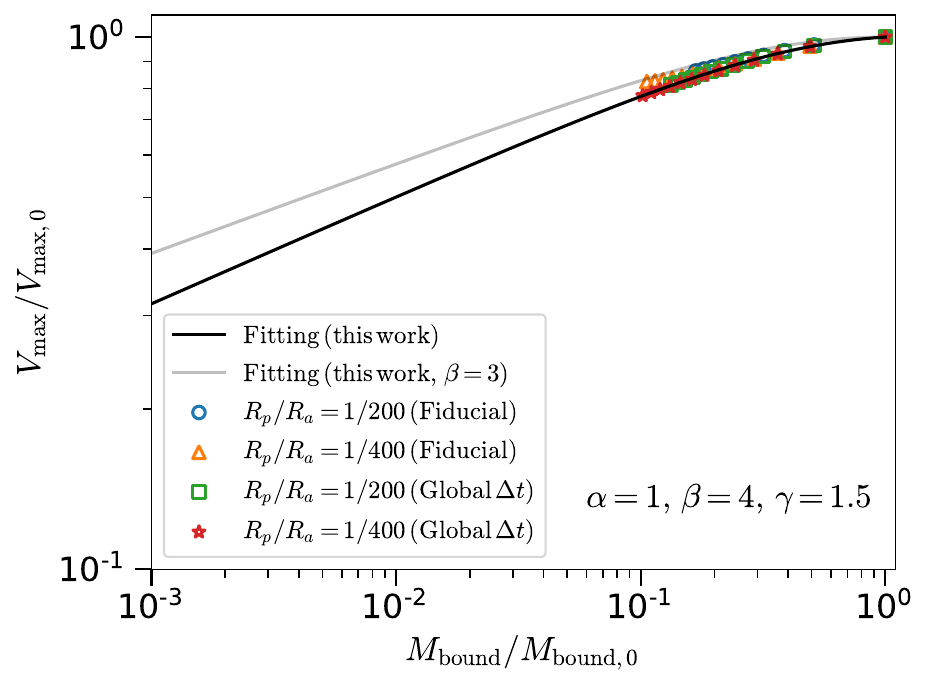}
\caption{$R_{\mathrm{max}}$ and $V_{\mathrm{max}}$ for $(\alpha,\beta)=(1,4)$ and different values of $\gamma$ from $N$-body simulations using {\sc Gadget-4}. Solid black lines show the fitting functions found in this work. For comparisons, the fitting functions for the same $(\alpha, \gamma)$, but $\beta=3$ are also shown in each panel (light gray lines).}
\label{fig:Rmax_Vmax_M_2}
\end{figure*}

\begin{figure*}
\includegraphics[width=0.9\columnwidth]{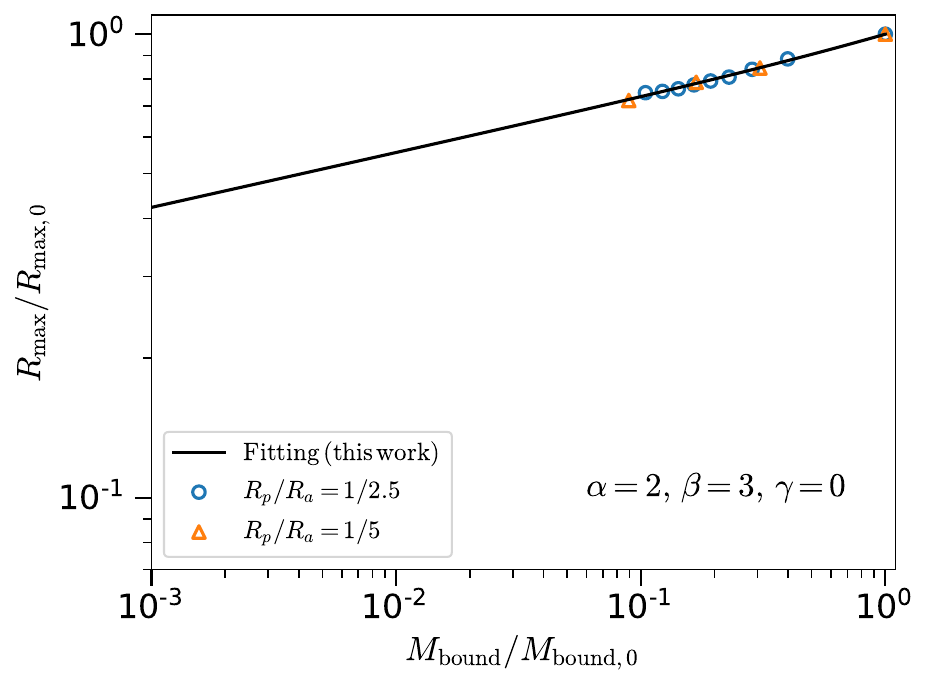}
\includegraphics[width=0.9\columnwidth]{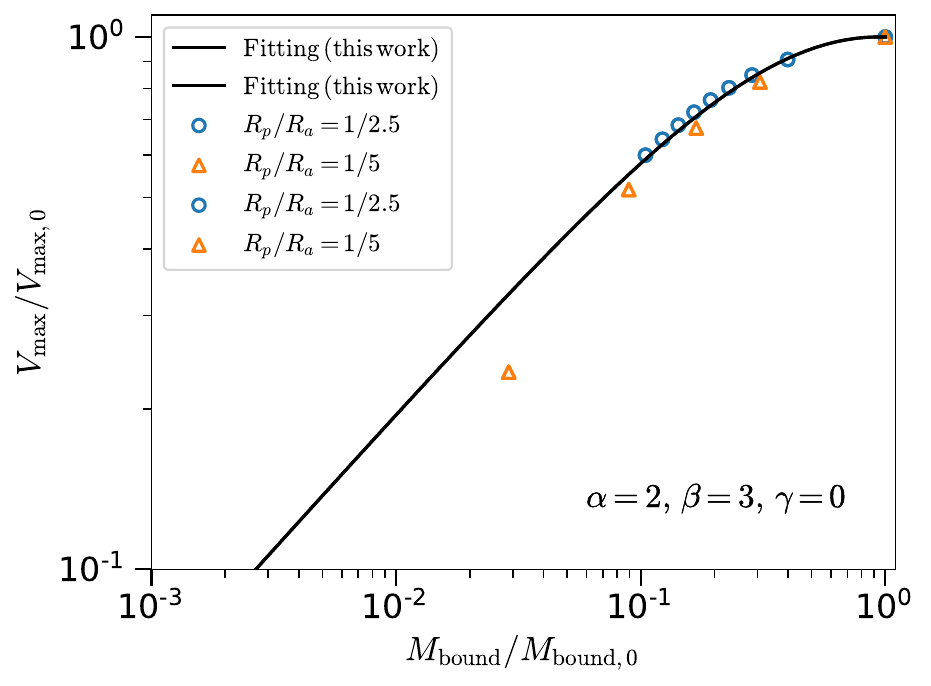}
\includegraphics[width=0.9\columnwidth]{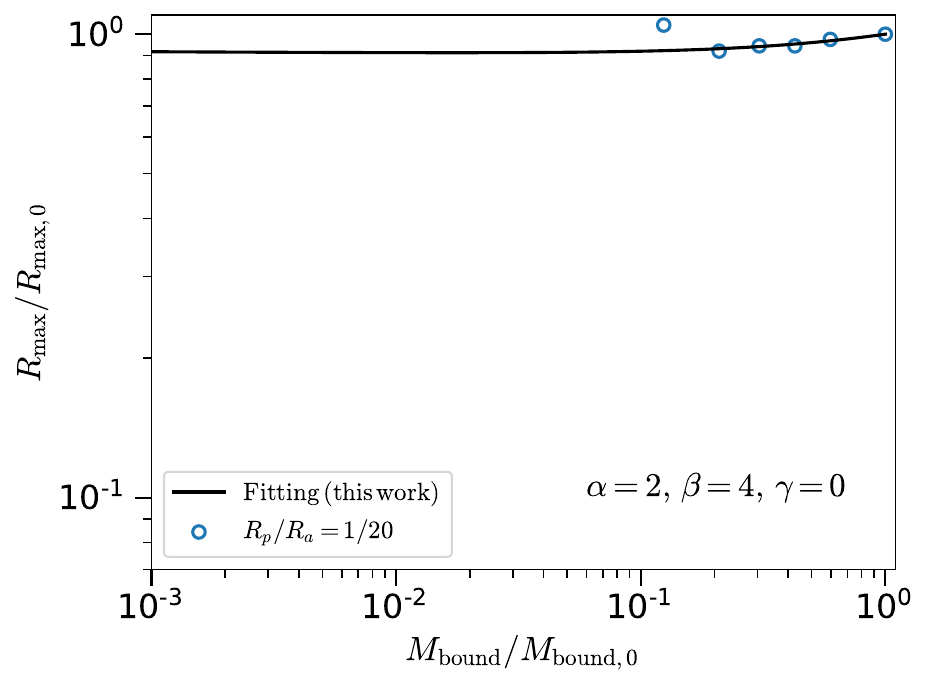}
\includegraphics[width=0.9\columnwidth]{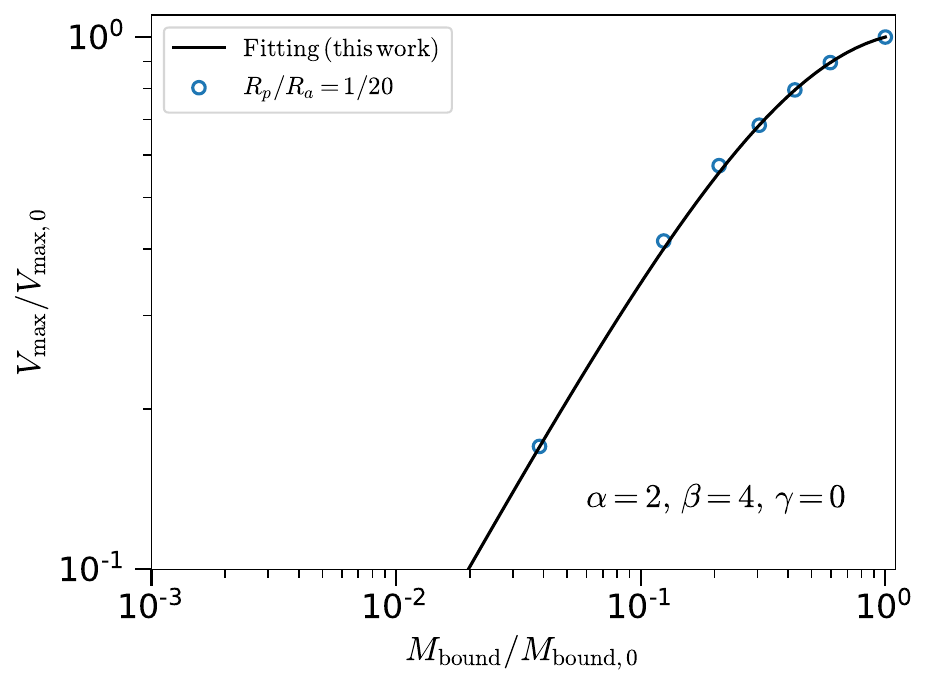}
\caption{$R_{\mathrm{max}}$ and $V_{\mathrm{max}}$ for $(\alpha,\gamma)=(2,0)$ and different values of $\beta$ from $N$-body simulations using {\sc Gadget-4}. Solid lines show the fitting functions found in this work.}
\label{fig:Rmax_Vmax_M_3}
\end{figure*}

In Sec.~\ref{sec:results}, we have shown $R_{\mathrm{max}}$ versus $V_{\mathrm{max}}$ tracks for different initial subhalo density profiles. In Figs.~\ref{fig:Rmax_Vmax_M_1}, \ref{fig:Rmax_Vmax_M_2} and \ref{fig:Rmax_Vmax_M_3}, we show $R_{\mathrm{max}}$ and $V_{\mathrm{max}}$ as functions of the bound mass fraction together with our best fit fitting formula.

\section{Fitting functions for density evolution}\label{appendix:den_fit_more}

\begin{figure*}
\includegraphics[width=0.9\columnwidth]{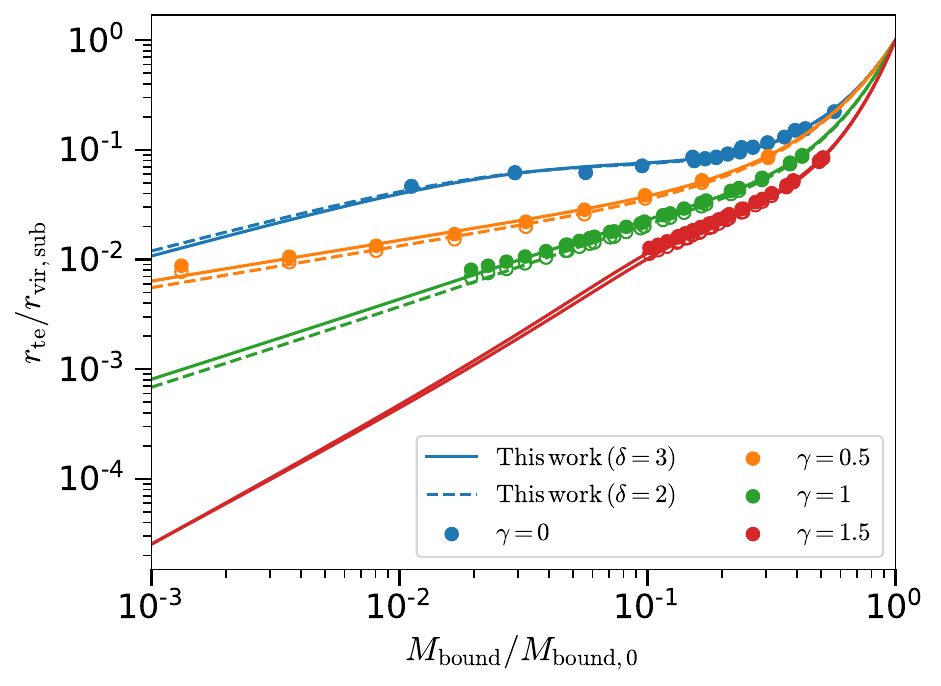}
\includegraphics[width=0.9\columnwidth]{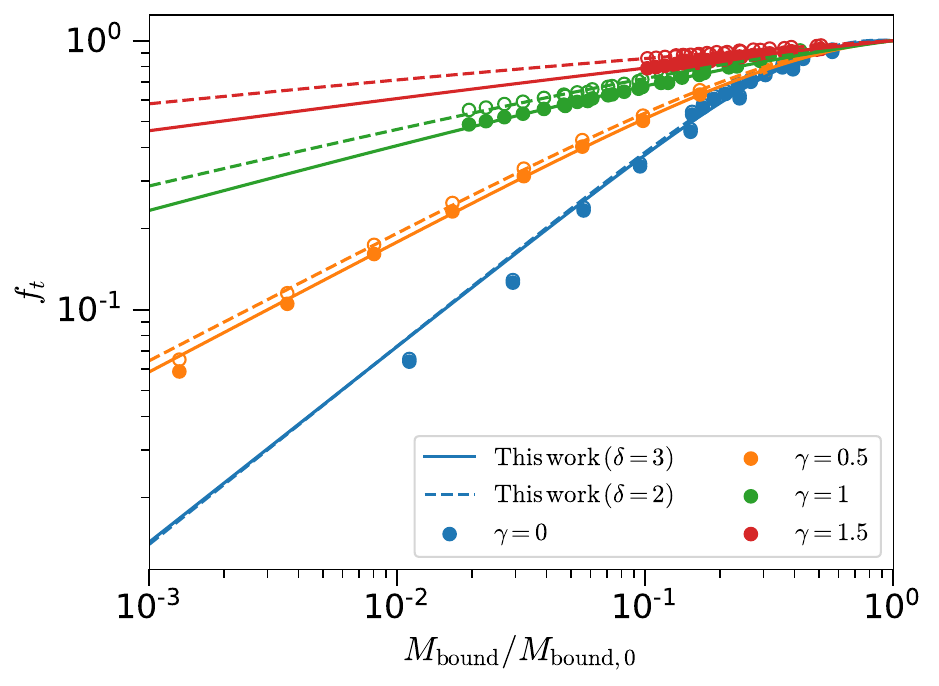}
\caption{Best-fit parameters for the density profile as a function of bound mass fraction for $(\alpha,\beta)=(1,4)$ and different $\gamma$. Left: the effect tidal radius in Eq.~\eqref{eq:transfer}. The filled (open) circles are obtained by fitting Eq.~\eqref{eq:transfer} to the density profiles measured from simulation assuming $\delta=3$ ($\delta=2$). Right: the normalization parameter $f_\mathrm{t}$ in Eq~\eqref{eq:transfer}.}
\label{fig:den_fit_more}
\end{figure*}

In Sec.~\ref{subsec:rho_evo}, we have shown the best fit parameters for the density transfer function for $(\alpha,\beta)=(1,3)$ and different $\gamma$. Best parameters for other combinations of $(\alpha,\beta,\gamma)$ are shown in Fig.~\ref{fig:den_fit_more}.

\begin{figure*}
\includegraphics[width=0.9\columnwidth]{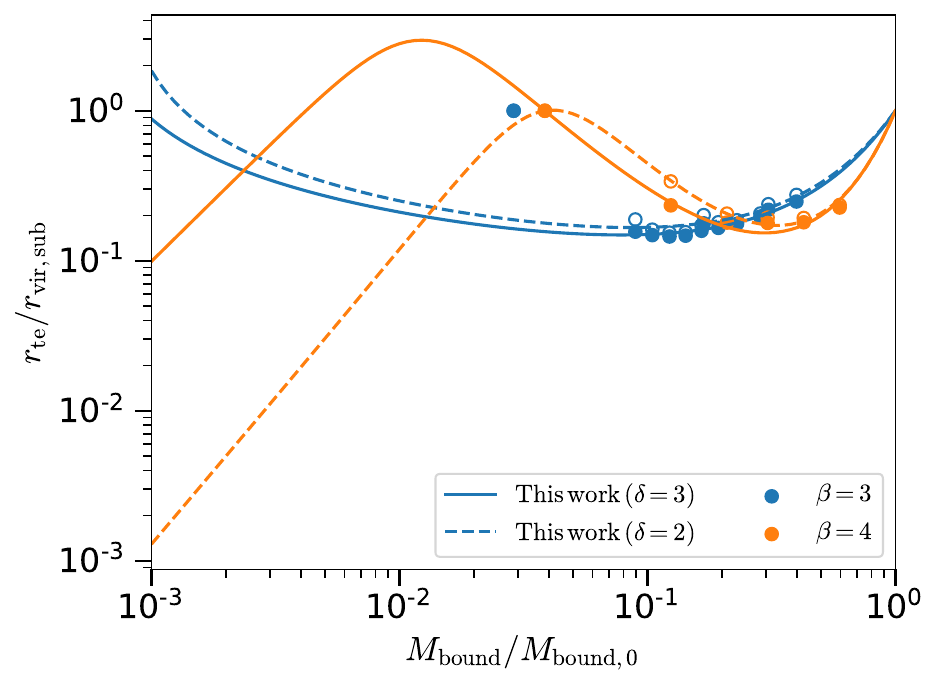}
\includegraphics[width=0.9\columnwidth]{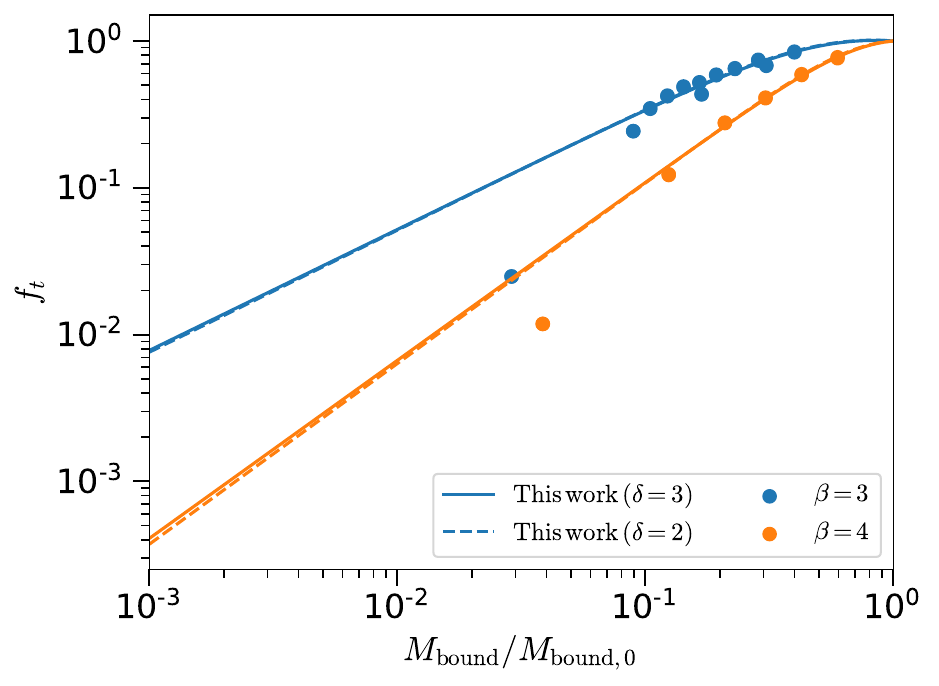}
\caption{Best-fit parameters for the density profile as a function of bound mass fraction for $(\alpha,\gamma)=(1,0)$ and different $\beta$. Left: the effect tidal radius in Eq.~\eqref{eq:transfer}. The filled (open) circles are obtained by fitting Eq.~\eqref{eq:transfer} to the density profiles measured from simulation assuming $\delta=3$ ($\delta=2$). Right: the normalization parameter $f_\mathrm{t}$ in Eq~\eqref{eq:transfer}.}
\label{fig:den_fit_more_2}
\end{figure*}

\section{Effects of subhalo concentration and time-evolving host potential}\label{appendix:diff_c}

\begin{figure*}
\includegraphics[width=0.9\columnwidth]{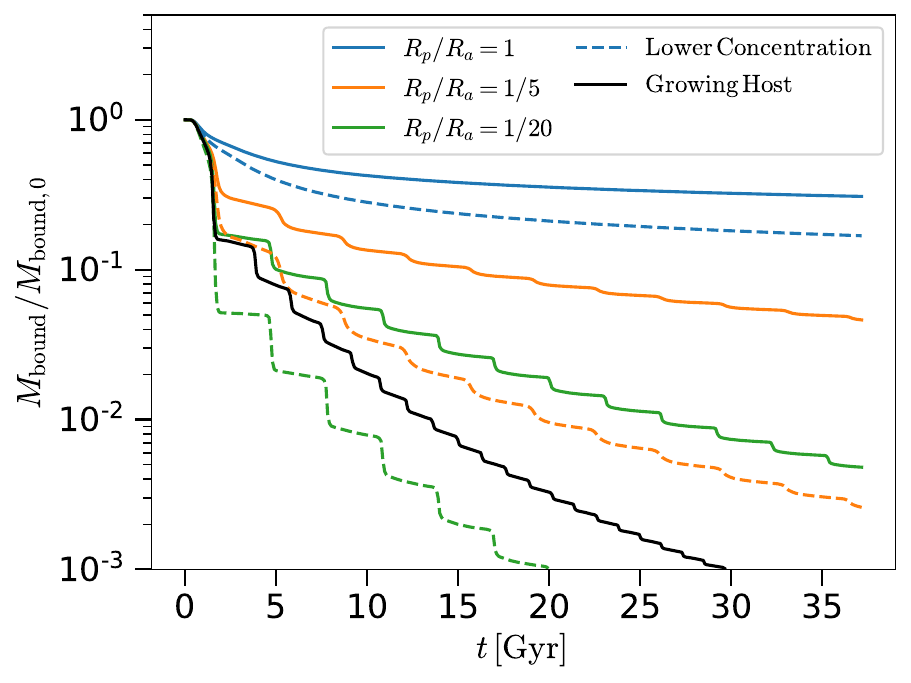}
\includegraphics[width=0.9\columnwidth]{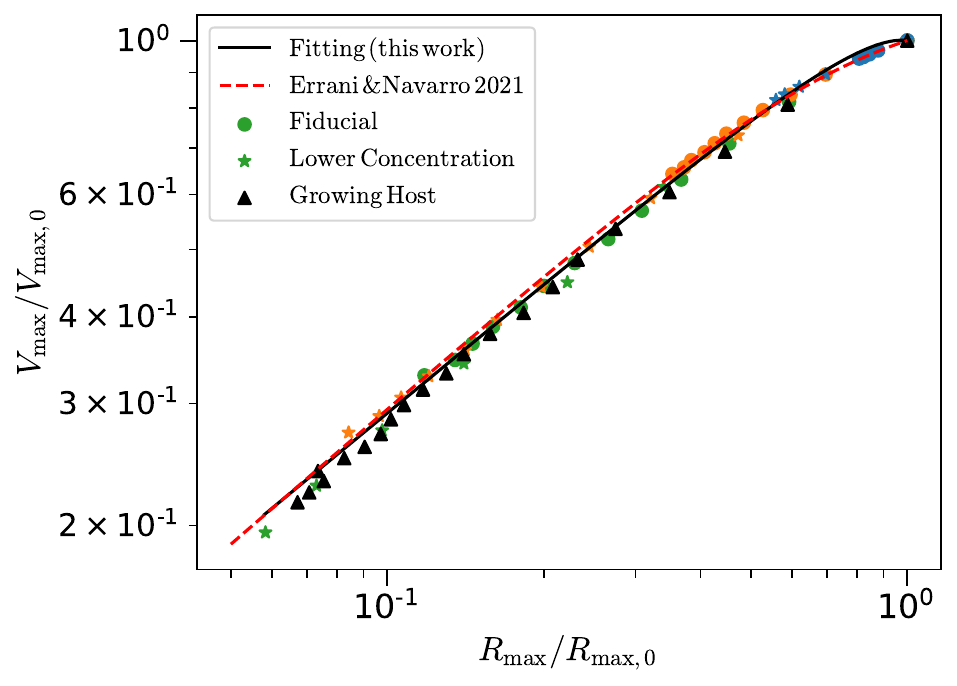}
\caption{Bound mass evolution and tidal tracks for NFW subhalos with different concentration (colored lines) and a time-evolving host potential (black line). For the case with evolving host potential, the subhalo is initially put on the same orbit as the static host case with $R_\mathrm{p}/R_\mathrm{a}=1/20$.}
\label{fig:tidal_track_diff_c_host}
\end{figure*}

To test how subhalo concentration may affect the tidal tracks. We run a few tests for subhalos with NFW profiles, i.e. $(\alpha,\beta,\gamma)=(1,3,1)$ and half of the fiducial concentration. As shown in Fig.~\ref{fig:tidal_track_diff_c_host}, subhalos with lower concentrations (colored dashed curves) are more influenced by the tidal stripping and have faster mass loss compare to the fiducial cases (colored solid curve). But the $R_{\mathrm{max}}$ versus $V_{\mathrm{max}}$ tracks are only marginally affected. A detailed study of an even larger change in subhalo concentrations as done in Green {\it et al.}~\cite{Green:2019zkz} is needed to determine the possible weak dependence of tidal tracks on subhalo concentrations.

In the fiducial simulations, we have a static host potential. However, in the realistic case, the host halo grows with time by accreting small halos. So we also run a test in which the host have a initial mass of $10^{12} \mathrm{M}_{\odot}$ (Milky Way size) and its mass grows linearly with time and reaches $8\times 10^{12} \mathrm{M}_{\odot}$ (group size) at the end of the simulation. The subhalo has an initial velocity that matches the static host case with $R_\mathrm{p}/R_\mathrm{a}=1/20$. Again, the subhalo has faster mass loss, but tidal tracks are only marginally affected, see the black curve (left panel) and triangles (right panel) in Fig.~\ref{fig:tidal_track_diff_c_host}.

\section{Effects of galactic disk and bulge}\label{appendix:disk}

In the fiducial simulations, we assume the host have an NFW profile and neglect any possible contribution from the baryons in the halo. To test whether our results are affected by the baryonic potential, we add a Miyamoto and Nagai disk potential~\cite{Miyamoto:1975}
\begin{equation}
\Phi_{\mathrm{MN}}=-\frac{G M_{\mathrm{MN}}}{\sqrt{x^2 + y^2 + (R_d+\sqrt{z_d^2+z^2})^2}},
\label{eq:MN_disk}
\end{equation}
and a Hernquist bulge potential~\cite{Hernquist:1990be}
\begin{equation}
\Phi_{\mathrm{HQ}}=-\frac{G M_{\mathrm{HQ}}}{R+A_{\mathrm{HQ}}},
\label{eq:HQ}
\end{equation}
to the static host potential. For the fiducial host mass of $10^{12} \mathrm{M}_{\odot}$, we choose the following parameters in Eqs.~\eqref{eq:MN_disk} and \eqref{eq:HQ} to mimic the Milky Way disk and bulge~\cite{McMillan:2011wd,Sameie:2018chj}:
\begin{eqnarray}
M_{\mathrm{MN}}&=&6.98 \times 10^{10} \mathrm{M}_{\odot},\\
R_d&=&6.0 \, {\mathrm{kpc}},\\
z_d&=&1.2 \, {\mathrm{kpc}},\\
M_{\mathrm{HQ}} &=& 1.05 \times 10^{10} \mathrm{M}_{\odot},\\
A_{\mathrm{HQ}}&=&0.46 \, {\mathrm{kpc}}.
\end{eqnarray}
The subhalo is assumed to have an NFW profile at the beginning of the simulation. We choose the ratio of pericenter to apocenter distances to be $R_\mathrm{p}/R_\mathrm{a}=1/50$ so that the subhalo can enter within the disk radius $R_d$. The subhalo orbital plane is tilted by an angle of $\pi/6$ with respect to the disk plane. 

\begin{figure*}
\includegraphics[width=0.9\columnwidth]{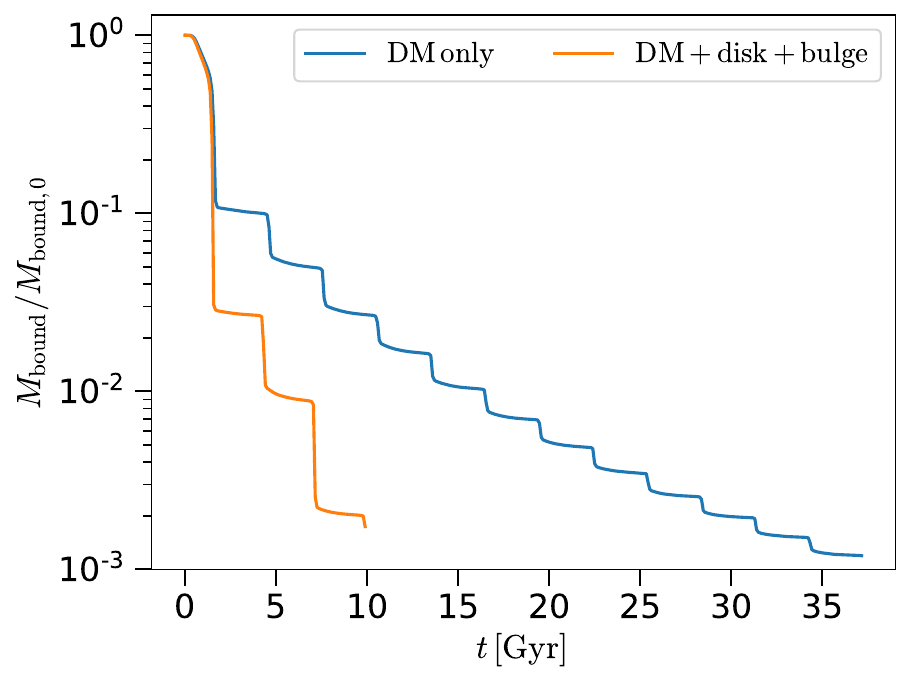}
\includegraphics[width=0.9\columnwidth]{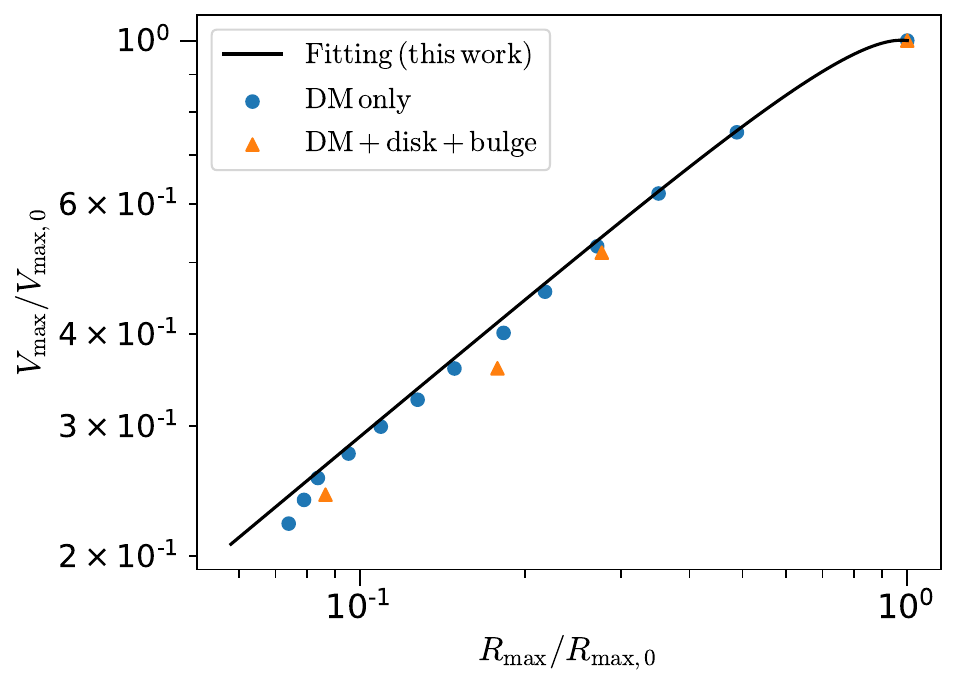}
\caption{Bound mass evolution and tidal tracks for NFW subalhos with and without the galactic disk and bugle potentials.}
\label{fig:tidal_track_with_disk}
\end{figure*}

As shown in Fig.~\ref{fig:tidal_track_with_disk}, the subhalo has a faster mass loss compared to the fiducial ``DM-only" simulation. But only a small difference (less than $10\%$) is observed in the $V_{\mathrm{max}}$ versus $R_{\mathrm{max}}$ track.

\clearpage
\bibliography{reference}



\end{document}